# Deep *Chandra* Observations of ESO 428-G014: II. Spectral Properties and Morphology of the Large-Scale Extended X-ray Emission

G. Fabbiano[a], A. Paggi[a, b,c,d], M. Karovska[a], M. Elvis[a], W. P. Maksym[a], G. Risaliti[e,f], Junfeng Wang[g]

*a. Harvard-Smithsonian Center for Astrophysics, 60 Garden St. Cambridge MA 02138, USA*
*b. Dipartimento di Fisica, Universita' degli Studi di Torino, via Pietro Giuria 1, I-10125 Torino, Italy*
*c. Istituto Nazionale di Fisica Nucleare, Sezione di Torino, via Pietro Giuria 1, 10125 Torino, Italy*
*d. INAF-Osservatorio Astrofisico di Torino, via Osservatorio 20, 10025 Pino Torinese, Italy*
*e. Dipartimento di Fisica e Astronomia, Università di Firenze, via G. Sansone 1, I-50019 Sesto Fiorentino (Firenze), Italy*
*f. INAF - Osservatorio Astrofisico di Arcetri, Largo E. Fermi 5, I-50125 Firenze, Italy*
*g. Department of Astronomy and Institute of Theoretical Physics and Astrophysics, Xiamen University, Xiamen, 361005, China*

Abstract

We present a deep *Chandra* spectral and spatial study of the kpc-scale diffuse X-ray emission of the Compton thick (CT) AGN ESO428-G014. The entire spectrum is best fit with composite photoionization + thermal models. The diffuse emission is more extended at the lower energies (<3 keV). The smaller extent of the hard continuum and Fe Kα profiles imply that the optically thicker clouds responsible for this scattering may be relatively more prevalent closer to the nucleus. These clouds must not prevent soft ionizing X-rays from the AGN escaping to larger radii, in order to have photoionized ISM at larger radii. This suggests that at smaller radii there may be a larger population of molecular clouds to scatter the hard X-rays, as in the Milky Way. The diffuse emission is also significantly extended in the cross-cone direction, where the AGN emission would be mostly obscured by the torus in the standard AGN model. Our results suggest that the transmission of the obscuring region in the cross-cone direction is ~10% than in the cone-direction. In the 0.3-1.5 keV band, the ratio of cross-cone to cone photons increases to ~84%, suggesting an additional soft diffuse emission component, disjoint from the AGN. This could be due to hot ISM trapped in the potential of the galaxy. The luminosity of this component ~5×10$^{38}$ erg s$^{-1}$ is roughly consistent with the thermal component suggested by the spectral fits in the 170-900 pc annulus.

# 1. Introduction

This paper is the second of our series on the Compton-thick Active Galactic Nucleus galaxy (CT AGN) ESO 428-G014 (also called IRAS 01745-2914, MCG-05-18-002), based on deep *Chandra* observations of the nuclear region. ESO 428-G014, a southern barred early-type spiral galaxy [SAB(r)], at a distance of ~23.3 Mpc (NED; scale=112 pc/arcsec) has a highly obscured Compton Thick (CT, $N_H > 10^{25}$ cm$^{-2}$) Seyfert Type 2 nucleus, with a high ratio of [OIII] λ5007 to hard 2-10 keV observed flux (Maiolino et al 1998). It also has the second highest [OIII] flux among CT AGNs after NGC1068 (Maiolino & Riecke 1995, Risaliti et al 1999), so it is expected to be the second intrinsically brightest CT AGN below 10 keV. This nucleus is associated with a curved 5''-long 6 cm. radio jet (Ulvestad & Wilson 1989) and with extended Hα and [OIII] optical line emission (Falcke et al 1996, Falcke, Wilson & Simpson 1998).

Based on the first ~30ks *Chandra*/ACIS observation of ESO 428-G014, Levenson et al (2006) reported soft extended X-ray emission that also follows the extended line emission region. This feature, observed in other Seyfert galaxies (e.g. Bianchi et al 2006, Wang et al **2011c,** Paggi et al 2012), suggests the presence of a bi-conical emission region photo-ionized by the Active Galactic Nucleus (AGN). Levenson et al (2006) detected a prominent Fe Kα line with EW=1.6 ± 0.5 keV over a hard flat continuum >3 keV, typical of a highly obscured AGN, as well as several emission lines in the soft spectrum. Based on a spectral fit of the entire spectrum, including with the best nuclear reflection model, they concluded that the extended soft emission is photoionized, ruling out a strong thermal component (proposed by David et al 2006).

With the cumulative *Chandra* exposure of 154.5 ks (5 times longer than that available to Levenson et al. 2006), we have taken a new look at the X-ray emission of ESO 428-G014, aimed at establishing the detailed spectral and morphological properties of the emission. Because ESO 428-G014 is a CT AGN, the X-ray luminosity of the nuclear source is heavily attenuated, and the central source is not piled-up in ACIS-S. This allows us to explore the circumnuclear region to the smallest sub-arcsecond radii allowed by the *Chandra* resolution. In Paper I (Fabbiano et al 2017), we reported the discovery of kiloparsec extended components in the 3-6 keV hard continuum and Fe Kα 6.4 keV line emission. This result challenged the established picture that the hard continuum and Fe Kα emission are confined to the Compton thick obscuring circumnuclear clouds in CT AGNs (~100 pc scale, Marinucci et al 2012, 2017). Paper I suggested that this hard and Fe Kα extended emission may be caused by scattering by the galaxy ISM of photons escaping the nuclear region in the direction of the ionization cone.

In this paper, we study the extended emission component of ESO 428-G014 in the entire 0.3-8 keV band observable with Chandra. We describe the observations and data reduction (Section 2), and report the results of a detailed spectral analysis of the nuclear and extended components of the X-ray emission of ESO 428-G014 (Section 3). Aided by these results, we study the large-scale morphology of the emission in different energy bands, from 0.3 to 8.0 keV (Section 4). We discuss our results in Section 5.

## 2. Data Reduction & Analysis

Table 1. summarizes the observations used in this paper. The data were inspected and merged as reported in Paper I, to produce a deeper data set for imaging analysis. As explained in Paper I, all the data sets were processed to enable sub-pixel analysis. The final full-band (0.3-8 keV) central region of the *Chandra* image of ESO 428-G014 from the merged dataset is shown in Fig. 1. This area is well within the 2MASS optical ellipse of the galaxy from NED; major and minor axes correspond to 15.4 kpc and 8.9 kpc respectively, for a scale of 112 pc/arcsec. A central region of extended circumnuclear emission is evident, elongated along the major axis of the galaxy. It extends ~33'' (~3.7 kpc) in the SE-NW direction of the ionization cone and radio jet (Ulvestad & Wilson 1989), and ~17" (1.9 kpc) in the SW-NE (cross-cone) direction.

Table 1.

| ObsID | Instrument | $T_{exp}$(ks) | PI | Date |
|---|---|---|---|---|
| 4866 | ACIS-S | 29.78 | Levenson | 2003-12-26 |
| 17030 | ACIS-S | 43.58 | Fabbiano | 2016-01-13 |
| 18745 | ACIS-S | 81.16 | Fabbiano | 2016-01-23 |

In what follows, we first analyze the integrated spectrum of the central emission region, we then use the results of this analysis to study the large-scale spatial properties of the emission of well-defined spectral features. A number of point-like sources, possibly X-ray binaries in ESO428-G014 are visible in Fig. 1. Since the focus of this study is the extended circumnuclear emission, we excluded from the spectral analysis and radial profiles the regions containing identifiable extra-nuclear point sources. Circular regions containing point-like sources were excluded from the data considered for spectral analysis and radial profiles of the extended emissions (see below). The excluded regions are marked in Fig. 1. These regions are retained in the images. Unless explicitly noted, we used CIAO 4.8 and CALBD 4.7.2 for our data preparation and analysis.

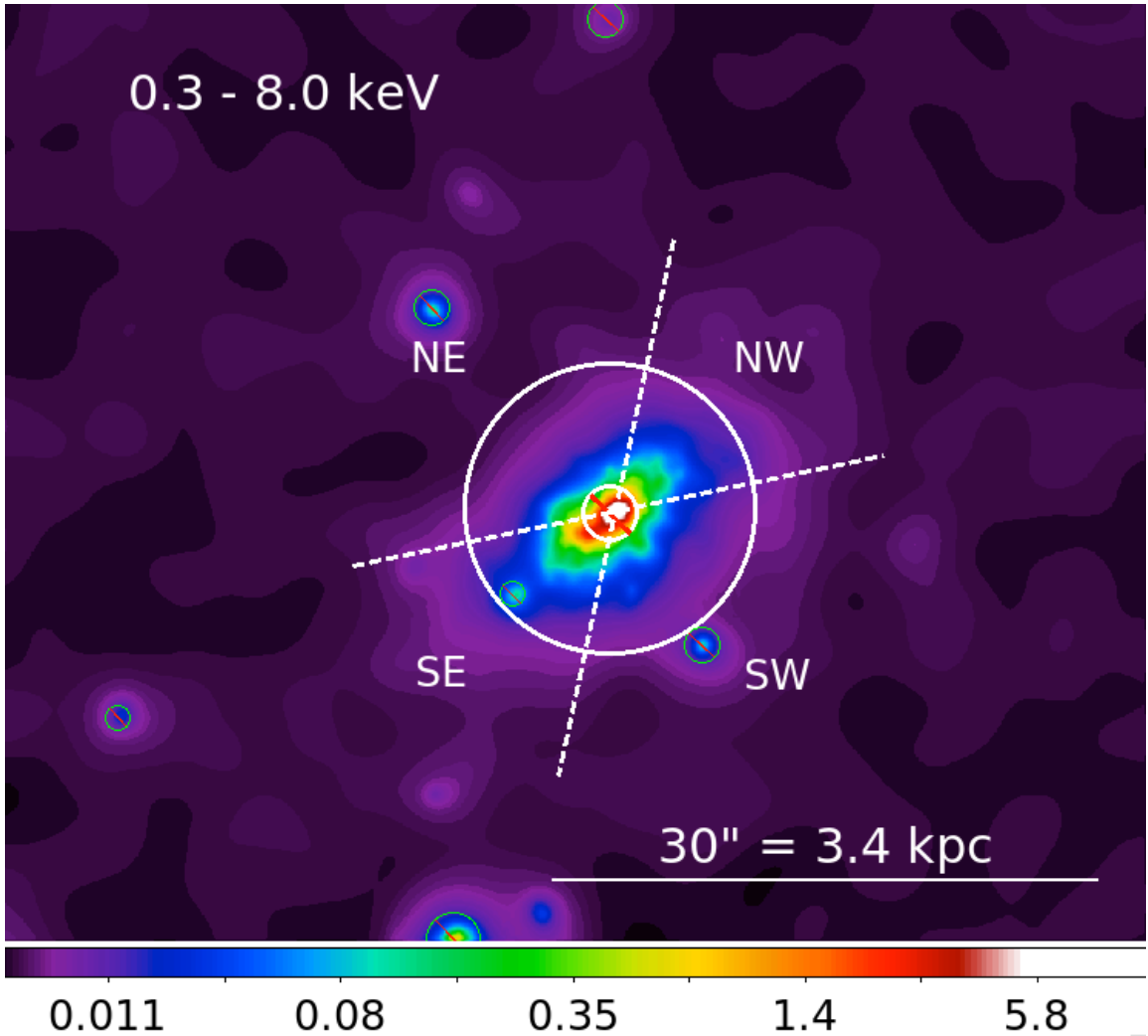

Fig. 1. Merged 0.3-8.0 keV *Chandra* ACIS image of ESO 428-G014, with image pixel = 1/8 ACIS pixel (corresponding to 0.06"), and adaptive Gaussian smoothing (1-25 pix scales, 10 counts under kernel, 30 iterations). The image scale and contours are logarithmic. The regions of non-nuclear point sources excluded from the spectral analysis and surface brightness profiles are marked. The logarithmic color scale is in number of counts per image pixel. The white circles denote the regions chosen for spectral analysis (Section 3), the dashed white lines define the cone (NW and SE) and cross-cone (NE and SW) sectors used for the radial profiles (Paper I; Section 4).

### 3. Spectral Properties of the Nuclear and Extended Emission

We extracted the spectrum for fitting from three different regions (see Fig. 1), all centered on J2000 RA = $7^h16^m31.2^s$, Dec = $-29°19'28.6"$: (1) A circle of 8'' radius centered on the nucleus, which encompasses the entire central emission region; this region provides the best comparison with the nuclear spectra obtained with other X-ray telescopes, which cannot discriminate between point-like and extended emission, lacking the sub-arcsecond spatial resolution of *Chandra*. **We include this fit for completeness and to facilitate future work.** (2) A circle of 1.5'' to encompass the nuclear emission proper, and the surrounding inner circumnuclear emission. (3) A 1.5''-8'' annulus, which includes the extended circumnuclear emission only. The background was extracted from larger source-free surrounding regions. We binned the spectra to have 20 counts/bin, and we fitted them to models in the 0.3-8.0 keV range.

### 3.1 Hard continuum and Fe Kα fit and empirical model of the soft line emission

Following Levenson et al. 2006, we first fitted the data with XSPEC (Arnaud 1996), using a multi component model. Since this nucleus is Compton thick, we modeled the hard continuum (E>4 keV) with purely reflected light using the PEXRAV model (Magdziarz & Zdziarski 1995) with power-law photon index $\Gamma=1.9$. To this we added several unresolved emission lines, leaving both energy and amplitude free to vary. A second power-law with $\Gamma=1.9$ was also needed (as found by Levenson et al 2006) to fit adequately the softer emission. We used only line of sight Galactic absorption to model the low-energy absorption.

Table 2 lists the results of this fit for the three extraction regions. The errors on line fluxes were calculated simultaneously, with multiple lines varying. As discussed in Paper I, continuum components are present in both the central 1.5'' radius circle and in the 1.5''- 8'' annulus. The Fe Kα emission is mostly associated with the nuclear region, although is also detected in the 1.5''-8'' annulus (see Paper I). The Fe XXV line is only detected in the spectra including the nuclear region.

Table 2.
**Spectral fitting of the central emission[a]**

| Region | Counts (error) | Norm. PEXRAV (ph cm$^{-2}$ s$^{-1}$) | Norm. 2$^{nd}$ Po. Law (ph cm$^{-2}$ s$^{-1}$) | $\chi_\nu^2$ | $\nu$ (d.o.f.) (Continuum + lines) |
|---|---|---|---|---|---|
| 8'' circle | 6983 (84) | $1.8\times10^{-3}$ | $4.0\times10^{-5}$ | 0.87 | 160 |
| 1.5'' circle | 4608 (68) | $1.5\times10^{-3}$ | $2.4\times10^{-5}$ | 0.76 | 122 |
| 1.5'' – 8'' annulus | 2375 (49) | $0.2\times10^{-3}$ | $2.0\times10^{-5}$ | 0.85 | 164 |

| *Emission lines:* | | |
|---|---|---|
| Energy (keV) (8'') (1.5'') (1.5''- 8'') | Flux ($10^{-6}$ ph cm$^{-2}$ s$^{-1}$) (8'') (1.5'') (1.5'' – 8'') | Identification |
| 0.58 +/- 0.01[b] | 64.5 +/- 10.4 | *OVII 7->1 |
| 0.58 +/- 0.01 | 39.8 +/- 8.2 | |
| 0.60 +/- 0.02 | 18.9 +/- 6.2 | |
| 0.73 +/- 0.01[b] | 29.0 +/- 3.4 | *Fe XVII 2->1 |
| 0.72 +/- 0.01 | 20.3 +/- 2.9 | |
| 0.75 +/- 0.01 | 9.3 +/- 2.0 | *Fe XVII 5->1 |
| 0.83 +/- 0.01 | 24.1 +/- 2.3 | *O VIII 19->1 (0.84), Fe XVII 27->1 (0.83) |
| 0.83 +/- 0.01 | 14.7 +/- 1.9 | |
| 0.85 +/- 0.03 | 8.1 +/- 1.6 | |
| 0.93 +/- 0.01[b] | 18.1 +/- 2.0 | *Ne IX  7->1 |
| 0.93 +/- 0.02 | 12.7 +/- 1.5 | |
| 0.93 +/- 0.02 | 4.3 +/- 1.4 | |
| | | |
| 1.03 +/- 0.01[b] | 11.1 +/- 1.2 | *Ne X  4->1 |
| 1.03 +/- 0.01 | 8.1 +/- 1.0 | |
| 1.03 +/- 0.02 | 3.4 +/- 0.8 | |
| | | |
| 1.12 +/- 0.01 | 5.3 +/- 1.0 | *Fe XXIII 15->1 |
| 1.12 +/- 0.01 | 3.8 +/- 0.9 | |
| 1.14 +/- 0.02 | 1.5 +/- 0.5 | |
| | | |
| 1.24 +/- 0.01[b] | 3.6 +/- 0.6 | *Fe XX  306->1 |
| 1.24 +/- 0.01 | 2.9 +/- 0.5 | |
| 1.29 +/- 0.03 | 0.8 +/- 0.5 | |
| | | |
| 1.34 +/- 0.02[b] | 2.7 +/- 0.5 | Mg XI  2->1 |
| 1.34 +/- 0.02 | 1.7 +/- 0.4 | |
| 1.37 +/- 0.04 | 0.64 +/- 0.54 | |
| | | |
| 1.47 +/- 0.02 | 1.1 +/- 0.4 | Mg XII  4->1 |
| 1.47 +/- 0.02 | 1.0 +/- 0.3 | |
| | | |
| 1.73 +/- 0.01[b] | 2.2 +/- 0.4 | Si K$\alpha$  2->1 |
| 1.73 +/- 0.02 | 1.0 +/- 0.3 | |
| 1.73 +/- 0.02 | 1.0 +/- 0.3 | |
| | | |
| 1.85 +/- 0.02[b] | 1.8 +/- 0.4 | Si XIII K$\alpha$  2->1 |
| 1.85 +/- 0.03 | 1.2 +/- 0.3 | |
| 1.82 +/- 0.02 | 0.8 +/- 0.3 | |
| | | |
| 2.32 +/- 0.02[b] | 1.4 +/- 0.4 | S K$\alpha$ 2->1 |
| 2.31 +/- 0.02 | 1.1 +/- 0.3 | |
| 2.41 +/- 0.03 | 0.5 +/- 0.2 | |
| | | |
| 6.40 +/- 0.01[b] | 8.9 +/- 0.8 | Fe K$\alpha$ 2->1 |
| 6.40 +/- 0.01 | 7.6 +/- 0.7 | |
| 6.37 +/- 0.02 | 1.6 +/- 0.4 | |
| | | |
| 6.71 +/- 0.03 | 2.0 +/- 0.5 | Fe XXV 7->1 |
| 6.70 +/- 0.04 | 1.6 +/- 0.5 | |

[a] Assumed line of sight $N_H=2.1\times10^{21}$ cm$^{-2}$, $\Gamma$ PEXRAV =1.9, $\Gamma$ 2$^{nd}$ power law = 1.9
[b] Emission lines also detected by Levenson et al (2006), using only ObsID 4866
*Lines blended in the ASCIS-S spectrum <1.3 keV. These are tentative identifications.

Emission lines at energies <3.0 keV are detected in both inner and outer regions. Of these lines, those at energies <1.3 keV are blended in our spectrum, so their detection has a certain degree of uncertainty. Some of these lines, however, e.g. O VII, Ne IX and Ne X are present in other AGN spectra (e.g., NGC 4151, Ogle et al 2000 with gratings, and Wang et al 2011c, partially blended in an ACIS spectrum). The Mg, Si, and S lines at energies 1.3-2.4 keV, even if partially blended, are relatively isolated and visible in the spectrum and residuals over the continuum, as is the Fe K$\alpha$ narrow line. For the 1.5'' circle, Fig. 2 shows the spectral data, model and residuals. The insert in the top panel of

Fig. 2 shows an enlargement of the hard spectrum, characterized by a featureless continuum plus the two Fe K emission lines. The Fe Kα emission is dominated by the neutral 6.4 keV line. The Fe XXV line, visible as the excess to the right of the narrow neutral line, accounts for ~1/5 or less of the line emission in the 6-7 keV band.

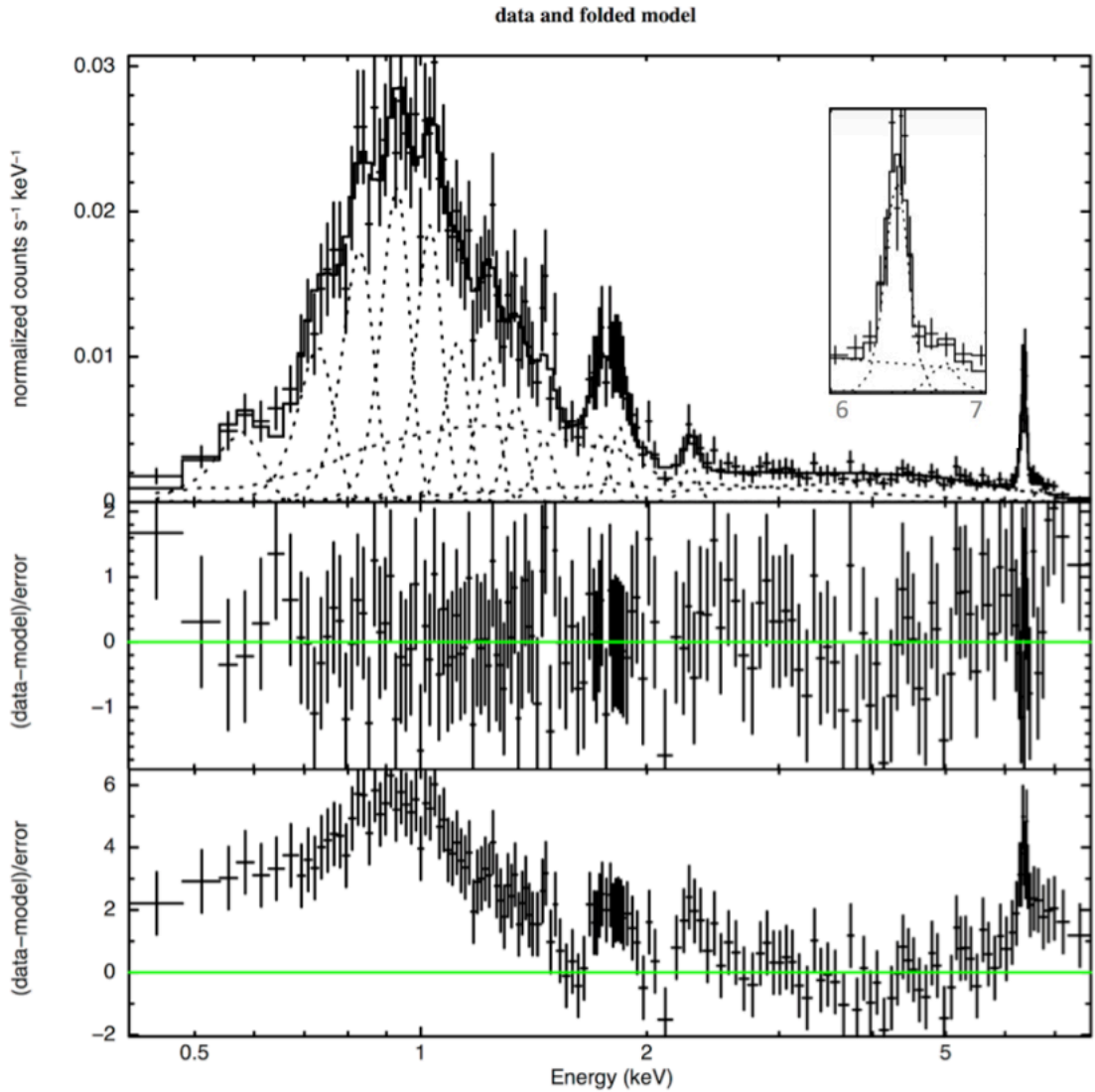

**Fig. 2** – Spectrum from the inner 1.5'' radius circle with best fit (top), best-fit residuals (middle) and residuals from best-fit continuum model (bottom). The insert in the top panel shows in detail the 6-7 keV spectrum, with the neutral and XXV Fe K lines best fits.

## 3.2 Physical models of the soft emission

To investigate possible physical emission mechanisms, we then fitted the spectra with photoionization (CLOUDY[1]; Ferland et al. 1998) and thermal (APEC[2]; Foster et al 2012) models, and combinations of these models. In all the fits, the model included the best-fit nuclear continuum and Fe Kα line model from Table 2. In all our fitting experiments, we started with one-component models and added additional components to improve the fit quality. We applied the F-test to estimate the improvement in the fit quality. We also inspected the best-fit residuals to identify cases when we obtain overall formally good fits (i.e. reduced $\chi^2$~1), but have correlated significant residuals in certain energy bands ($\chi^2$ is not sensitive to correlated residuals.) We used these inspections to justify the addition of a new component to the model (see examples in Fig. 3).

We fitted the data from the three circular regions used in Section 3.1. Moreover, for the 1.5''-8'' annulus, which includes only the extended diffuse emission, we also fitted the spectra from the cone and cross–cone directions (as defined in Paper I, see Fig. 1), to explore possible azimuthal dependencies of the emission parameters. None was found, within statistics.

### 3.2.1 Photoionization

For photoionization (CLOUDY) only models, the F-test shows a significant improvement is obtained by increasing the number of model components to three in the 8'' circle and in the nuclear region (1.5'') circle. In the 1.5"-8" annulus, while a significant improvement occurs going from one to two components, adding the third component produces only marginal improvement (see Table 3). Adding a fourth photo-ionization component does not improve the fit, and slightly increases the $\chi^2$ in the extended and annulus regions. In the nuclear region, the reduced $\chi^2$ goes from 0.94 to 0.93, with a F-test probability of 0.69, indicating that the 4th component is not required. In Table 4 we list the best-fit parameters for the 3-components photo-ionization emission for the spectra from the three regions.

Visual inspection of the fit results shows that even a 3-component photoionization model – which gives an overall good reduced $\chi^2$ - cannot account satisfactorily for the spectrum in the entire energy range. Fig. 3 (top left) shows an example of how the 3-component photoionization model fails to fit the 8" circle region below 1 keV.

### 3.2.2 Thermal

Levenson et al (2006) found that single temperature thermal spectra gave unacceptable fits to the extended soft emission with reduced $\chi^2 > 3$, failing in reproducing the narrow peaks observed around 0.8, 0.9, and 1.1 keV. Using APEC models with solar abundances, we find that the reduced $\chi^2$ tend to be larger than for the photoionization fits with the same number of components (Table 3). Fig. 3 (top-right) shows that the 3-component thermal model fails to fit the spectral range between 1.5 and 3 keV.

---

[1] http://www.nublado.org/
[2] http://www.atomdb.org/

Table 3.
Reduced $\chi^2$, d.o.f. and F-test P for Photoionization and Thermal models

|  | Photoionization | | Thermal | |
| --- | --- | --- | --- | --- |
|  | $\chi_\nu^2$ ($\nu$) | F-Test P | $\chi_\nu^2$ ($\nu$) | F-Test P |
| (8" circle) | | | | |
| 1-comp. | 3.6 (186) | … | 4.3 (187) | … |
| 2-comp. | 1.4 (183) | $5.8\times10^{-38}$ | 2.3 (185) | $8.6\times10^{-24}$ |
| 3-comp. | 1.0 (180) | $3.9\times10^{-13}$ | 1.3 (183) | $5.5\times10^{-28}$ |
| (1.5" circle) | | | | |
| 1-comp. | 3.4 (147) | … | 2.8 (148) | … |
| 2-comp. | 1.3 (144) | $8.3\times10^{-31}$ | 1.3 (146) | $3.3\times10^{-24}$ |
| 3-comp. | 1.0 (141) | $1.3\times10^{-10}$ | 1.1 (144) | $1.1\times10^{-7}$ |
| (1.5"-8" annulus) | | | | |
| 1-comp. | 2.2 (86) | … | 4.3 (87) | … |
| 2-comp. | 1.1 (83) | $1.0\times10^{-13}$ | 1.4 (85) | $1.9\times10^{-22}$ |
| 3-comp. | 0.9 (80) | $1.5\times10^{-3}$ | 1.2 (83) | $2.4\times10^{-4}$ |

Table 4.
Best-Fit Parameters with 3-Components Models*

| Photoionization | | Thermal (APEC with solar abundances) | |
| --- | --- | --- | --- |
| (8'' circle) | | | |
| log(U1) = -0.9 ± 0.1 | log(NH1) > 22.4 | kT1 = 0.14 ±0.01 | EM1 = 10.4 ±2.6×$10^{-5}$ |
| log(U2) = 1.93 ± 0.05 | log(NH2) < 23.2 | kT2 = 0.79 ±0.02 | EM2 = 4.7 ±0.2 × $10^{-5}$ |
| log(U3) = 1.0 ± 0.1 | log(NH3) = 19.5 ± 0.3 | kT3 = 3.9 ±0.5 | EM3 = 11.9 ±0.7 × $10^{-5}$ |
| (1.5'' circle) | | | |
| log(U1) = -1.2 ± 0.3 | log(NH1) > 22.2 | kT1 = 0.15 ±0.02 | EM1 = 6.0 ±0.2 × $10^{-5}$ |
| log(U2) = 1.9 ± 0.1 | log(NH2) unconstrained | kT2 = 0.80 ±0.03 | EM2 = 3.0 ±0.2 × $10^{-5}$ |
| log(U3) = 0.9 ± 0.1 | log(NH3) = 19.7 ± 0.3 | kT3 = 2.5 ±0.3 | EM3 = 5.9 ±0.5×$10^{-5}$ |
| (1.5''- 8'' annulus) | | | |
| log(U1) = -1.0 ± 0.6 | log(NH1) = 22.6 ± 0.5 | kT1 = 0.75 ±0.04 | EM1 = 1.6 ±0.1 × $10^{-5}$ |
| log(U2) > 1.8 | log(NH2) = 19.7 (unconstr.) | kT2 = 0.14 ±0.02 | EM2 = 4.5 ±2.0 × $10^{-5}$ |
| log(U3) = 1.0 ± 0.2 | log(NH3) < 19.9 | kT3 = 9.8 ±4.2 | EM3 = 6.2 ±0.6 × $10^{-5}$ |

* See Table 3 for fit statistics
U is the ionization parameter of each component
NH is the column density (cm$^{-2}$) – 'unconstrained' is when it cannot be constrained in the fitting interval between $10^{19}$ and $10^{23.5}$ cm$^{-2}$
kT is the temperature (keV)
EM is the normalization of the APEC model (cm$^{-5}$), which is proportional to the emission measure.

### 3.2.3 Mixed Photoionization and Thermal Models

To investigate models that would give a better fit over the entire spectral range (see Fig. 3), we tried a set of mixed photoionization and thermal models (in addition to the nuclear continuum model). These include an increasing number of photoionization components (from 1 to 3) plus a thermal component, and an increasing number of thermal components **(from 1 to 3)** plus a photoionization component. We also tried a 2-photoionization + 2-thermal model. The $\chi^2$, degrees of freedom, and formal F-Test probabilities are listed in Table 5 **(note that the 2-component and 2+2 component models occur twice in Table 5, for reasons of symmetry).** Best-fit parameters for chosen models are given in Table 6.

The 2 + 1 models produced a significant improvement of fit statistics over the 1+1 models (Table 5), however, even if the fit is formally acceptable, the 2 + 1 models still cannot fit well the entire spectral range. For example, the 2 photoionization + 1 thermal (APEC with solar abundances) model has an overall $\chi^2$ ~1 (Table 5), so the fit is formally acceptable; but, while this model does a good job at energies below 1 keV, it still produces significant residuals in the 1-2 keV band (not shown).

Adding a fourth thermal APEC component to the 3-component photoionization model eliminates both soft and hard residuals producing a good fit in the entire spectral range (Fig. 3, bottom left), and so does a composite model with 2 photoionization + 2 thermal component (Fig. 3, bottom right). In this case we obtain slightly smaller values of $\chi_\nu^2$ with respect to the models with 3-photoionization+1-thermal components, with a larger number of degrees of freedom. A 3-thermal +1-photoionization model also fits well the entire spectral range. In the central 1.5'' circle, models with either 3 photoionization components + a thermal component with kT~0.4 keV, or 2 thermal+ 2 photoionization components are preferred. In the latter case, the two temperatures would be 0.7 and 1.2 keV respectively.

In summary, we find $\chi^2 \leq 1$ for models with more than 2 components, so formally all these fits are acceptable (Tables 3 and 5). However, *the presence of coordinated residuals in definite energy ranges* (see e.g. Fig. 3**)** leads us to favor mixed photoionization-thermal models over photoionization-only (or thermal-only) multi-components models. The data do not allow us to favor a particular model over another, so the range of possible models will be considered in our discussion of the physical constraints on the emitting regions posed by our data (Section 5).

### 3.2.4 Replacing PEXRAV with XILLVER

PEXRAV, though still widely used, is a relatively simple model and numerous improved, more physical, models have been introduced in the subsequent two decades. To investigate whether these refinements affect our results, we have also performed spectral fits using the reflection code XILLVER (Garcia et al 2013), which allows for an ionized reflecting medium, and includes line production from this medium. The results for photoionization and thermal fits reported above do not change. The same best-fit components are derived as in Tables 3-6, and all are significant. We conclude that our constraints on photoionization and thermal components are robust to the choice of reflection model. The main difference is that there is no need to include an additional power-law in the model to obtain a good fit. This is true for all the spectral extraction regions. However, the more limited statistics in the 1.5''-8'' annulus make the result ill defined for this region.

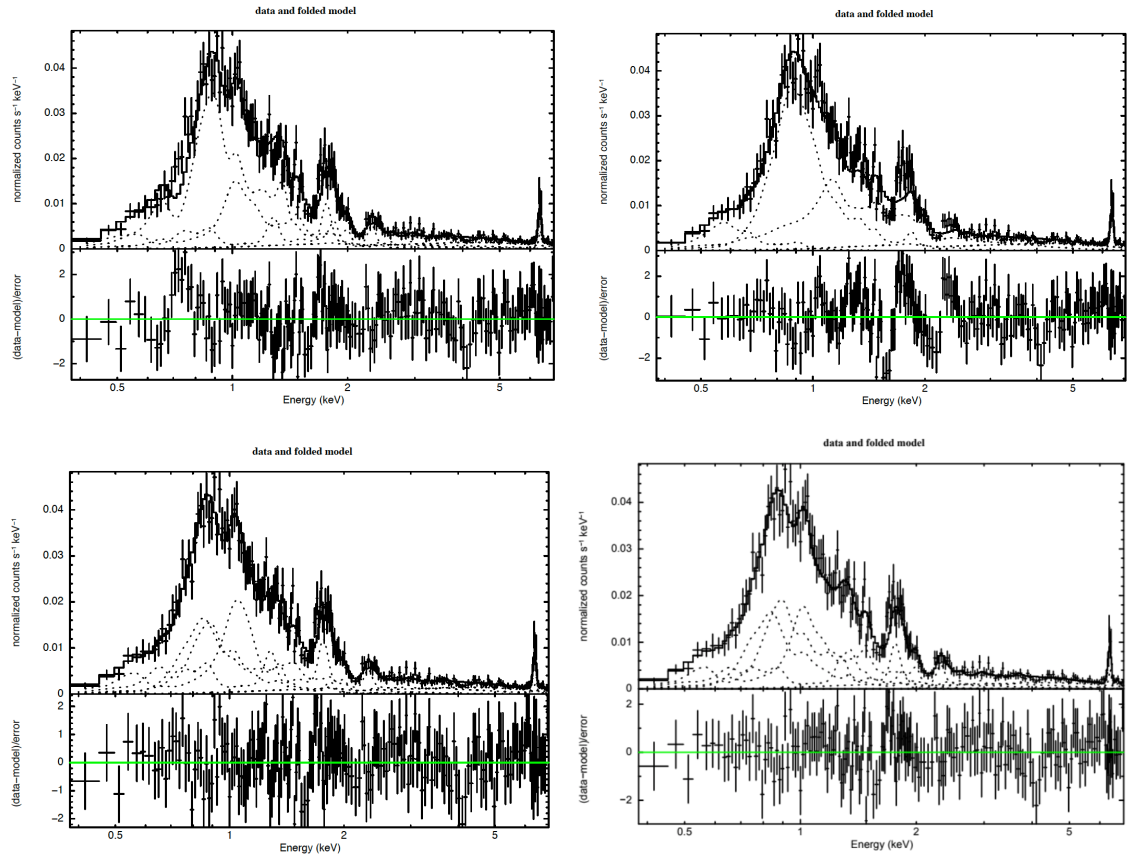

Fig. 3 – Top: Spectrum from the 8'' circle fitted with the nuclear PEXRAV component plus 3-component photoionization (left) and 3-component thermal (right). Note how the 3-photoionization model fails to fit the region below 1 keV, while the 3-thermal model fails to fit the region between 1.5 and 3 keV. Bottom: two examples of composite model fit: PEXRAV plus 3-component photoionization plus a thermal component model (left) and 2-photoionization + 2-thermal (right). The model best-fit components are shown in the upper panels of each figure as dashed lines. The lower panels show the fit residuals.

Table 5.
Reduced $\chi^2$, d.o.f. and F-test P for Mixed Photoionization and Thermal models

| | N-Photoionization + Thermal | | N-Thermal + Photoionization | | 4-Component* |
|---|---|---|---|---|---|
| | $\chi_v^2$ (ν) | F-Test P | $\chi_v^2$ (ν) | F-Test P | F-Test P |

(8" circle)
| | | | | | |
|---|---|---|---|---|---|
| 2-comp. | 1.3 (184) | … | 1.3 (184) | … | |
| 3-comp. | 1.0 (181) | $3.8 \times 10^{-10}$ | 1.0 (182) | $8.0 \times 10^{-12}$ | |
| 4-comp. | 0.9 (178) | $1.1 \times 10^{-5}$ | 0.8 (180) | $1.4 \times 10^{-5}$ | NA |

(1.5" circle)
| | | | | | |
|---|---|---|---|---|---|
| 2-comp. | 1.2 (145) | … | 1.2 (145) | … | |
| 3-comp. | 0.9 (142) | $1.3 \times 10^{-8}$ | 1.0 (143) | $9.0 \times 10^{-6}$ | |
| 4-comp. | 0.8 (139) | $9.2 \times 10^{-3}$ | 1.0 (141) | NA | $2.7 \times 10^{-6}$ |
| 2+2-comp.** | 0.8 (140) | *** | 0.8 (140) | $2.5 \times 10^{-8}$ | |

(1.5"-8" annulus)
| | | | | | |
|---|---|---|---|---|---|
| 2-comp. | 1.0 (84) | … | 1.0 (84) | … | |
| 3-comp. | 0.9 (81) | $4.6 \times 10^{-4}$ | 0.9 (82) | $3.7 \times 10^{-4}$ | |
| 4-comp. | 0.8 (78) | $1.7 \times 10^{-1}$ | 0.8 (80) | $4.5 \times 10^{-2}$ | NA |

\* F-Test comparing the 3-Photoionization+1-Thermal model fit to the 3-Thermal+1-Photoionization
\*\*The 2+2 component model is compared with the 3 components models
\*\*\* The F-test returns a very small P<$1. \times 10^{-12}$, but we are in a $\chi_v^2 <1$ regime
NA denotes cases where the F-test fails because $\chi_v^2$ increases with an additional component

Table 6.
Best-Fit Parameters with Mixed Models*

| 3-Components Photoionization + Thermal** | | 3-Components Thermal + Photoionization | |
|---|---|---|---|

(8'' circle)
| | | | |
|---|---|---|---|
| log(U1) = -0.8 ± 0.3 | log(NH1) > 22.3 | kT1 = 0.8 ±0.1 | EM1 = 3.5 ±0.5 × $10^{-5}$ |
| log(U2) = 1.9 ± 0.1 | log(NH2) = 19.3 (unconstr.) | kT2 = 0.18 ±0.03 | EM2 = 4.9 ±0.1 × $10^{-5}$ |
| log(U3) = 1.0 ± 0.1 | log(NH3) = 20.0 ± 0.6 | kT3 = 1.6 ±0.1 | EM3 = 5.4 ±0.6 × $10^{-5}$ |
| kT = 0.6 ± 0.1 keV | EM = 2.3 ± 0.6 × $10^{-5}$ | log(U)=-1.4±0.2 | log(NH)=22.6±0.3 |

(1.5'' circle)
| | | | |
|---|---|---|---|
| log(U1) = -1.0 ± 0.7 | log(NH1) > 21.7 | kT1 = 1.2 ±0.2 | EM1 = 1.9 ±0.6 × $10^{-5}$ |
| log(U2) = 1.9 ± 0.1 | log(NH2) = 19.6 (unconstr.) | kT2 = 0.34 ±0.04 | EM2 = 2.5 ±0.6 × $10^{-5}$ |
| log(U3) = 1.0 ± 0.2 | log(NH3) = 20.2 ± 0.7 | kT3 <20 | EM3 = 3.5 ±1.1 × $10^{-5}$ |
| kT = 0.4 +/- 0.1 keV | EM = 1.8 +/- 0.7 × $10^{-5}$ | log(U)=-1.0±0.1 | log(NH)=20.7±0.7 |
| log(U1) = -0.9 ±0.2 | log(NH1) > 21.9 | | |
| log(U2) = 0.8 ±0.2 | log(NH2) < 20.2 | | |
| kT = 0.7 ± 0.1 | EM = 1.4 ± 0.5 × $10^{-5}$ | | |
| kT = 1.5 ±0.2 | EM = 3.2 ±0.5 × $10^{-5}$ | | |

(1.5''- 8'' annulus)
| | | | |
|---|---|---|---|
| log(U1) = -1.0 ±0.1 | log(NH1) = 22.7 +/- 0.4 | kT1 = 0.8 ±0.1 | EM1 = 1.1 ±0.3 × $10^{-5}$ |
| log(U2) = 2.0 ± 0.1 | log(NH2) = 19.7 (unconstr.) | kT2 = 0.2 ±0.1 | EM2 = 1.0 ±0.7 × $10^{-5}$ |
| log(U3) = 0.9 ± 0.5 | log(NH3) < 21.6 | kT3 = 1.5 ±0.4 | EM3 = 1.5 ±0.5 × $10^{-5}$ |
| kT = 0.7 ± 0.2 keV | EM = 0.8 ± 0.5 × $10^{-5}$ | log(U)=-1.2±0.1 | log(NH)=22.3±0.3 |

\* See Table 5 for fit statistics; See Table 4 for description of parameters.
\*\* For the 1.5" circle we also display the results for the 2-photoionization+2-thermal fit.

## 4. The Extended Emission in Different Energy Bands

With *Chandra* we can also explore the nature of the X-ray emission by imaging the different spectral components (see Paper I). For this analysis we produced images in selected spectral 'slices' and we also explored the radial dependence of these emissions in azimuthal sectors. The images and radial profiles were built from 1/8 sub-pixel data to effectively oversample the Chandra PSF and overcome the limitations of the ACIS instrumental pixel, which is larger than the aimpoint mirror PSF. This method was used in and validated by our previous work on AGN circumnuclear regions, where we have demonstrated how the features recovered from the ACIS images match well with both higher instrument resolution *Chandra* HRC images, optical emission line clouds resoved with *HST*, and high-resolution radio features (e.g. Wang et al 2011a, b, c; Paggi et al 2012).

### 4.1 Images

To produce images, we processed the data with both smoothing and deconvolution algorithms. We used CIAO 4.8 tools, and the display application DS9 (http://ds9.si.edu/site/Home.html), which is also packaged with CIAO. Images produced with adaptive smoothing all reveal large-scale fainter low-surface brightness regions, and high surface brightness inner regions (see e.g. Fig. 1). To highlight some of the features, we then applied the Expectation through Markov Chain Monte Carlo (EMC2, Esch et al. 2004; Karovska et al. 2005, 2007) PSF-deconvolution. The EMC2 images in finer energy bands are shown in Figs. 4, 5, and 6 (details in figure captions). These deconvolved images are a reconstruction of the true shape of the source. Note that the soft band images (Fig. 4) cover twice the field as the hard band images (Figs. 5 and 6).

The emission is more extended at the lower energies, although elongated diffuse surface brightness can be seen at energies greater than 3 keV in the hard continuum (Fig. 5; as reported in Paper I). Figure 4 also shows that in the inner ~300pc (~3'') there is a markedly brighter extension to the SE in the emission line dominated 0.3-3.0 keV range. This extension is present throughout this energy band. This elongation is also present, but less prominent in the 3-6 keV continuum, at least up to 5 keV (Fig. 5). The deconvolved 5-6 keV image instead (Fig. 6) is dominated by the nuclear point source and suggests a round region of diffuse emission around the nucleus. The deconvolved Fe K$\alpha$ image (6.1-6.6 keV) is extended in the cone direction (Fig. 6). The overall morphology of the Fe K$\alpha$ emission is similar to that reported for the hard continuum in Paper I. The cross in Figs. 4, 5, and 6 is at the position of the center of the spectral extraction regions (see also Fig. 1), and given the data and deconvolution uncertainties is consistent with the hard nuclear source.

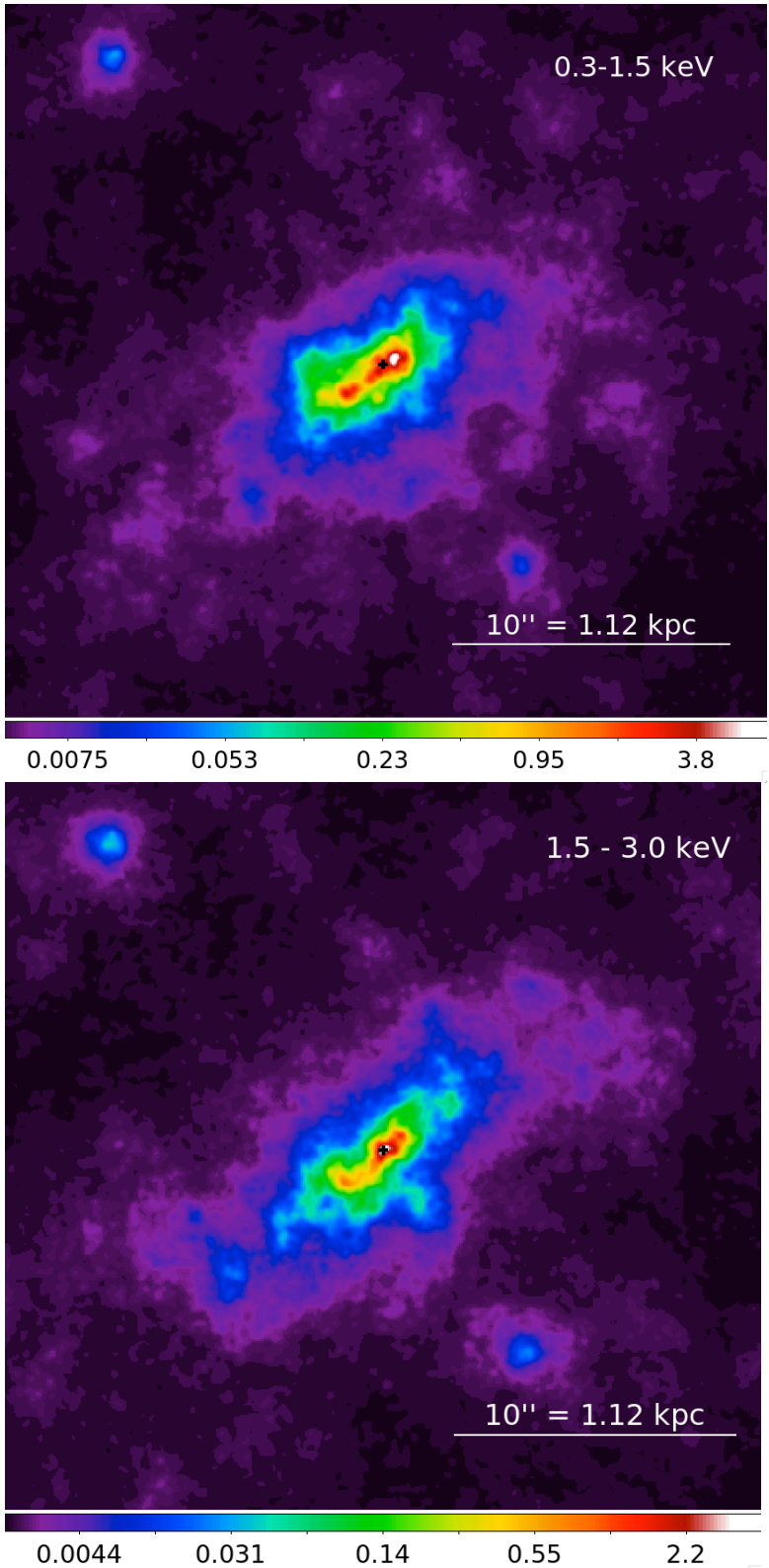

Fig. 4 - EMC2 PSF deconvolution images in the indicated energy bands (smoothed with 3pixel Gaussian). Color scale is in number of counts per image pixel. N is to the top and E to the left. The black cross marks the center of the 4-5 keV nucleus.

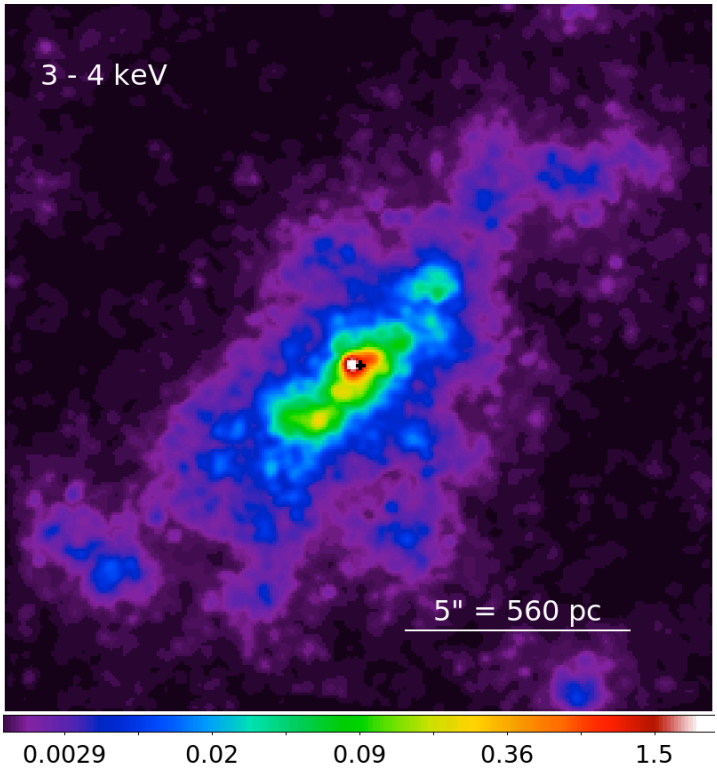
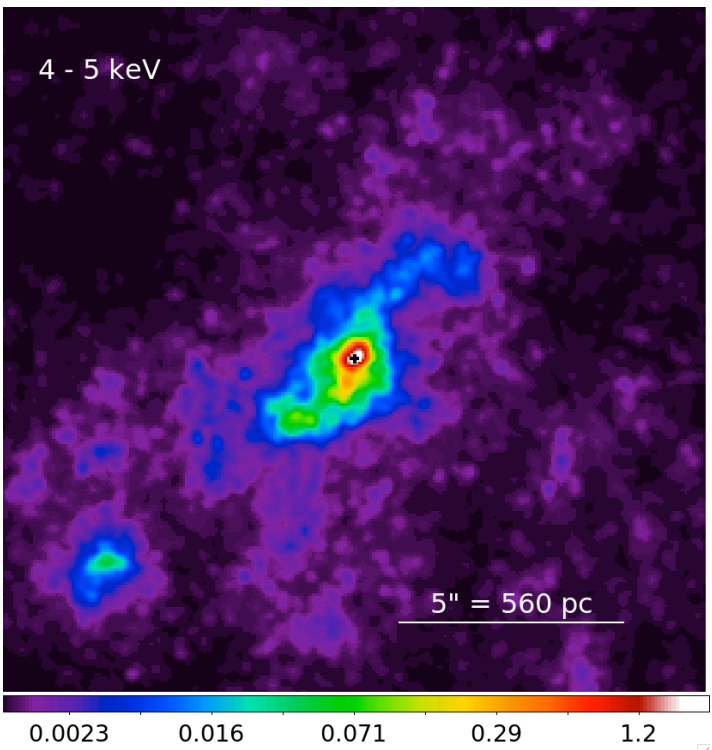

Fig. 5 – EMC2 PSF deconvolution images in the indicated energy bands (adaptive smoothing; see also Paper I for the 3.0-6.0 keV image). N is to the top and E to the left. The black cross marks the center of the 4-5 keV nucleus.

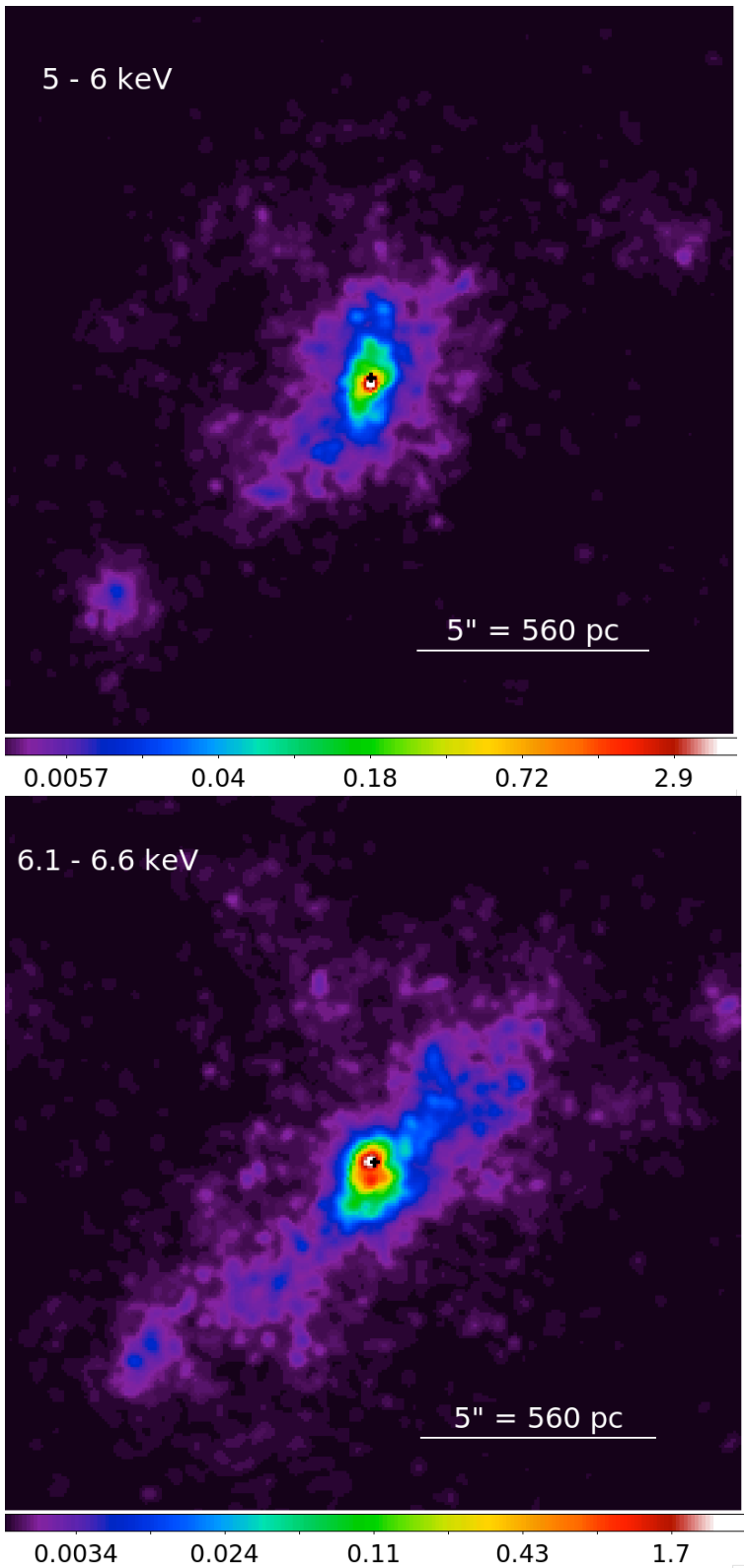

Fig. 6 – EMC2 PSF deconvolution images in the indicated energy bands (adaptive smoothing; see also Paper I for the 3.0-6.0 keV image). N is to the top and E to the left. The black cross marks the center of the 4-5 keV nucleus.

## 4.2 Radial Profiles

We followed the procedure outlined in Paper I to derive radial surface brightness profiles in different energy bands and different azimuthal sectors. The azimuthal sectors used in Paper I for the 3-6 keV continuum are also well suited for the study of the surface brightness profiles in other energy bands. Using the centroid of the spectral extraction, which is consistent with the position of the hard nuclear point source, we selected the SE cone between 101 and 173 degrees (measured counter-clockwise from North) and the NW cone between 285 and 349 degrees. Inspection of the images (see Section 4.1) shows that the surface brightness is more extended in the NW-SE direction (the 'cone') than in the NE-SW direction ('cross-cone'). The selection of angles was based on an azimuthal projection of the surface brightness.

To quantify the magnitude and significance of the extended components in different energy bands, we have compared a set of radial profiles from those energy bands, with a set of *Chandra* Point Source Function (PSF) models generated with CHART[3] and MARX[4] for the emission centroid position and the same energy bands. We used images with sub-pixel binning of 1/8 for generating these profiles. Off-nuclear point sources in each image were in all cases subtracted from the radial profiles, but their count contribution is in any case small (<3%). The profiles plots show the absolute background-subtracted surface brightness value of the data (in black), in counts/pixel as measured from the images in each energy band. The background values were estimated from large off-source areas from each energy band image and subtracted from the data. The PSF profiles (in red) were generated in the same energy bands as the data and normalized to the total counts within the central 0.5'' radius. Since there are changes in the central surface brightness distribution in different energy bands, this normalization approach is valid for comparing each sector to the relevant PSF, but in the softer energy band (0.3 – 1.5 keV) we can see an inner point that has a larger surface brightness than the central bin (see Fig. 8). This reflects the fact that the nucleus is heavily absorbed in the soft band and therefore the central bin may not correspond to the highest surface brightness value (compare Fig. 4 with Figs. 5, 6). The bin size was chosen to contain a minimum of 10 counts.

As shown in Fig. 7, the softer 0.3-1.5 keV band is the most extended. Comparison with the PSF profiles (in red) shows that the surface brightness is extended in both cone and cross-cone directions, although the extent in the cross-cone direction is smaller. In the cone direction, the surface brightness can be traced out to ~15'' (~1.7 kpc) radially, for a total extent of ~2.4 kpc. In the cross-cone direction, the radial extent is ~10'' (1.1 kpc). The 1.5-3.0 keV profile can be traced only out to ~10'' (1.1 kpc) in the cone direction, and ~5'' (~560 pc) cross-cone, in excess of the respective PSF profiles.

As also shown by the images (Figs. 4 - 6), the SE cone profiles show higher surface brightness relative to the NW cone profiles, particularly in the inner ~5" (Fig. 7). This inner region is where radio continuum and optical line emission are also present (Falcke et al 1996, 1998, Paper I). We will explore more fully this regions in Paper III (in preparation). Past 5'' the SE and NW profiles follow a similar radial trend.

---

[3] URL: http://cxc.harvard.edu/ciao/PSFs/chart2/
[4] URL: http://cxc.harvard.edu/ciao/threads/marx/

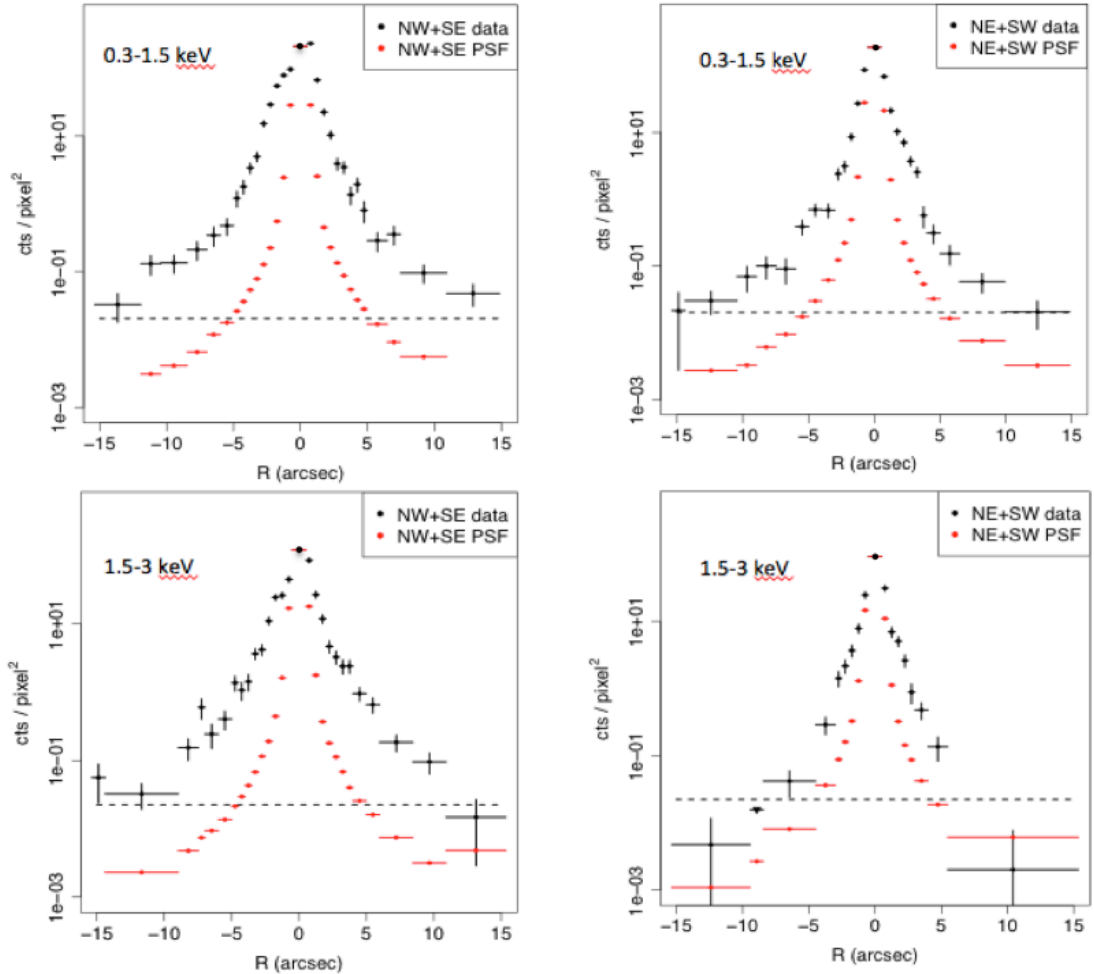

Fig. 7 – Background-subtracted radial profiles of the emission in the cone (NW+SE, left, data plotted from SE to NW) and cross-cone sectors (NE+SW, right, data plotted from SW to NE), in the indicated energy bands. 1/8 pixel re-binning was used. The PSF in the same energy band was normalized to the source image in the central 0.5'' circle. The bin size (shown) was chosen to contain a minimum of 10 counts. Errors are 1σ. The dashed horizontal line is plotted to indicate the level of the field background for each plot, derived from the same image as the profiles. Note that, since the profiles are background subtracted, points below the background level are valid data. Outside the plotted range, the profiles are noise-dominated.

Figure 8 shows the cone and cross-cone profiles for the 3-6 keV band, dominated by the continuum spectral component (Section 3), and incremental 1-keV-wide bands within this interval. As reported in Paper I, the 3-6 keV surface brightness extends radially out to ~1kpc from the nucleus in the cone direction, and therefore this continuum emission is not directly attributable to the emission from the immediate vicinities of the central active nucleus. The extent of the surface brightness decreases at higher energies and in the cross-cone direction, but in all cases, the emission is more extended than that of a PSF in the same energy band, notably at ~5''. As already suggested by Fig. 6, the cone and cross-cone profiles are similar in the 5-6 keV band.

Figure 9 (upper panels) shows the profiles in the Fe K$\alpha$ band. As reported in Paper I, the emission is significantly extended in the cone direction, while it is close to the PSF in the cross-cone. The profiles in the 7-8 keV continuum band are shown in the lower panels of Fig. 9; these are both consistent with the PSF.

In summary, radial profiles and images (Section 4.1) show kiloparsec-scale emission both at the energies <3 keV, which are dominated in X-rays by line emission (Section 2), and in the 3-6 keV continuum and Fe K$\alpha$ line that have usually been interpreted as connected strictly with the nuclear emission of the AGN. While this extent is more pronounced in the 'cone' direction, which is the direction of the ionization cone and the radio jet, the emission is extended also in the 'cross-cone' direction, at all energies up to ~5 keV. There is a general trend for the outer diffuse emission being generally less extended with increasing energy, which we discuss below. Above 6 keV, the emission is consistent with a nuclear point source.

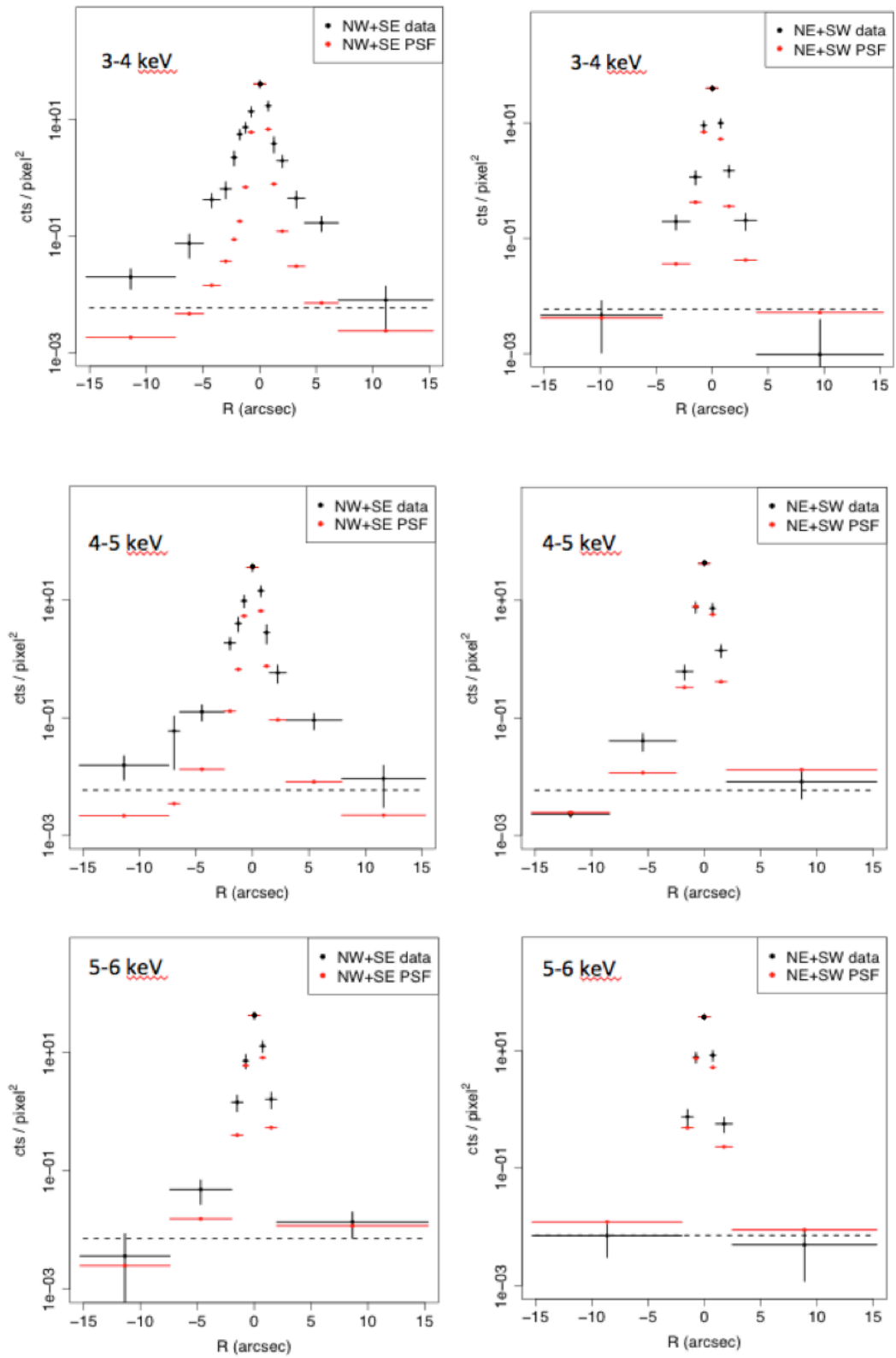

Fig. 8 - Same as Fig. 7, for the indicated energy bands.

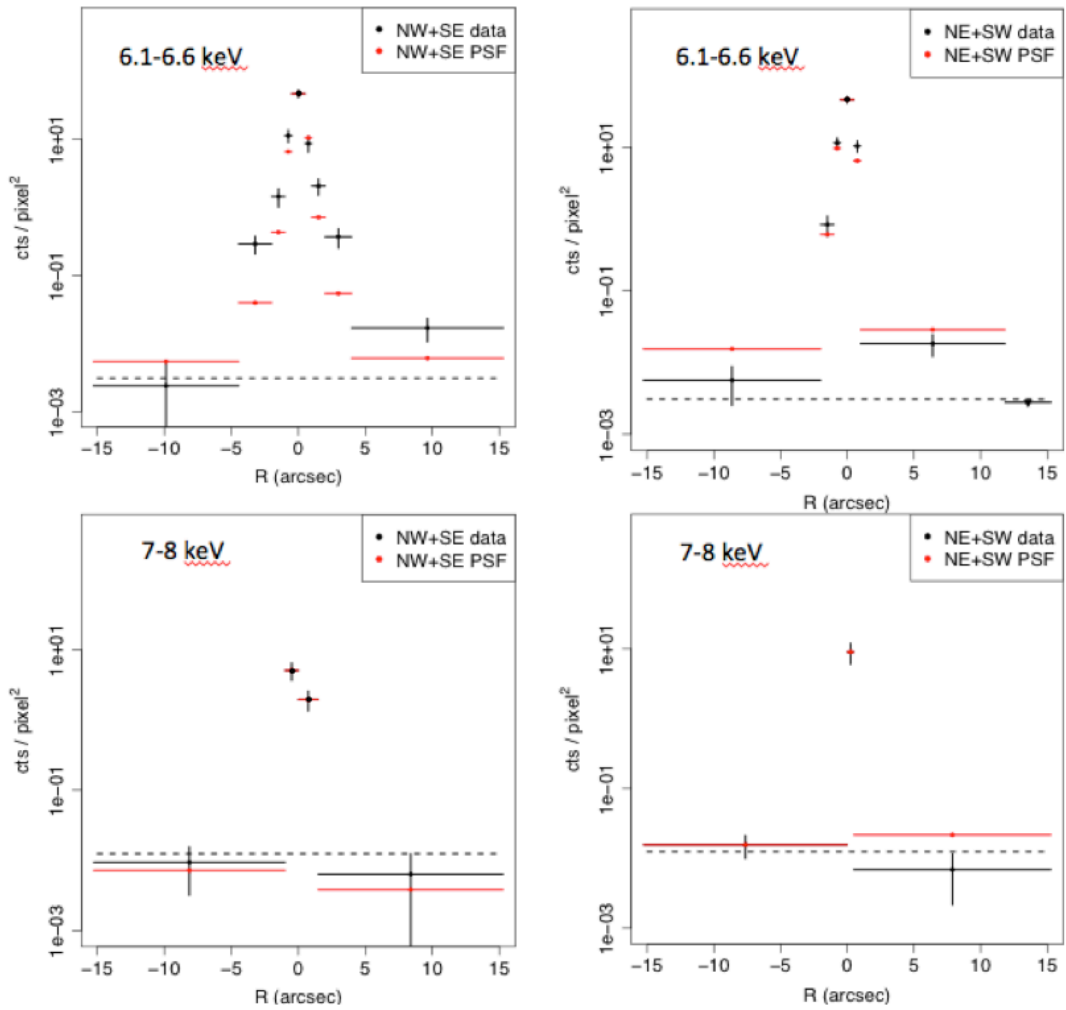

Fig. 9 – Same as Fig. 7, for the 6.1-6.6 keV band, dominated by the Fe Kα neutral line. (top), and the 7-8 keV band (bottom).

## 4.3 Large-Scale Emission at Different Energies: Extent and Intensity of 'Cone' and 'Cross-Cone' Emission

To measure the extent of the X-ray diffuse emission, minimizing signal to noise effects that may make it more visible at larger radii in the energy bands where the overall emission is more intense, we calculated the full-width-half-maximum (FWHM) in log as a proxy of the extent. This FWHM was measured from a spline approximation of the profiles and corresponds to the total extent of emission at surface brightness 1% of peak brightness. Fig. 10 plots this FWHM as function of the energy, for the profiles where significant extent over a point source distribution can be measured. Although it is not a measure of the total detected extent of the large-scale emission, this FWHM is a reliable comparison of the effective extent as a function of energy, because at this surface brightness the radial profiles are of good statistical significance. The error bars in Fig. 10 are indicative mostly of the uncertainty deriving from the bin sizes under question, which in turn reflect the surface brightness statistics in our adaptive binning (see Section 4). These errors were evaluated with a Monte-Carlo simulation and are equivalent to $1\sigma$. Because of the relatively low signal to noise, a single bin was used for the 4 – 6 keV interval. Given that the measured FWHM are typically a few arcseconds, these measurements are not significantly affected by the *Chandra* PSF (FWHM~1/3").

Fig. 10 shows a trend of decreasing FWHM with increasing energy both for the cone and the cross-cone directions. In the cone direction, we measure ~0.9 kpc at the lowest energies, decreasing to ~0.6 kpc at ~4 keV. While in the cross-cone direction there is continuous decrease of the FWHM, in the cone the FWHM is constant up to ~3keV, given the uncertainties in our measurements. The lowest energy bin (0.3-1.5 keV) and the 4-6 keV bin differ at the $3.6\sigma$ level in the cone direction and $5\sigma$ level in the cross-cone. The Fe K$\alpha$ profile in the cone direction is extended (as established in Paper I), while it is consistent with the PSF in the cross-cone (therefore this point is missing from Fig. 10). Similarly, we do not plot the 7-8 keV points because the emission is consistent with the PSF.

To measure the full extent of the diffuse emission in our observations, we also measured the width at which the background-subtracted surface brightness from the radial profiles is consistent with the background surface brightness in the same energy band. This width corresponds to ~$10^{-4}$ of the peak surface brightness and is significantly uncertain (see error bars in Fig. 10). We find that the maximum extent at the low energies in the cone direction is ~3.5 kpc (>2.5 kpc at $1\sigma$). The total detected extent in the Fe K$\alpha$ line is ~2.4 kpc (>1.8 kpc at $1\sigma$). In the cross-cone direction, there is a sharp decrease of total maximum detected extent between the 0.3-1.5 keV and the 1.5-3.0 keV bands. The cross-cone extent at the higher energies appears to be around ~1.9 kpc, although this may be affected by the poor resolution (increasing bin sizes) in the faintest low-surface brightness regions.

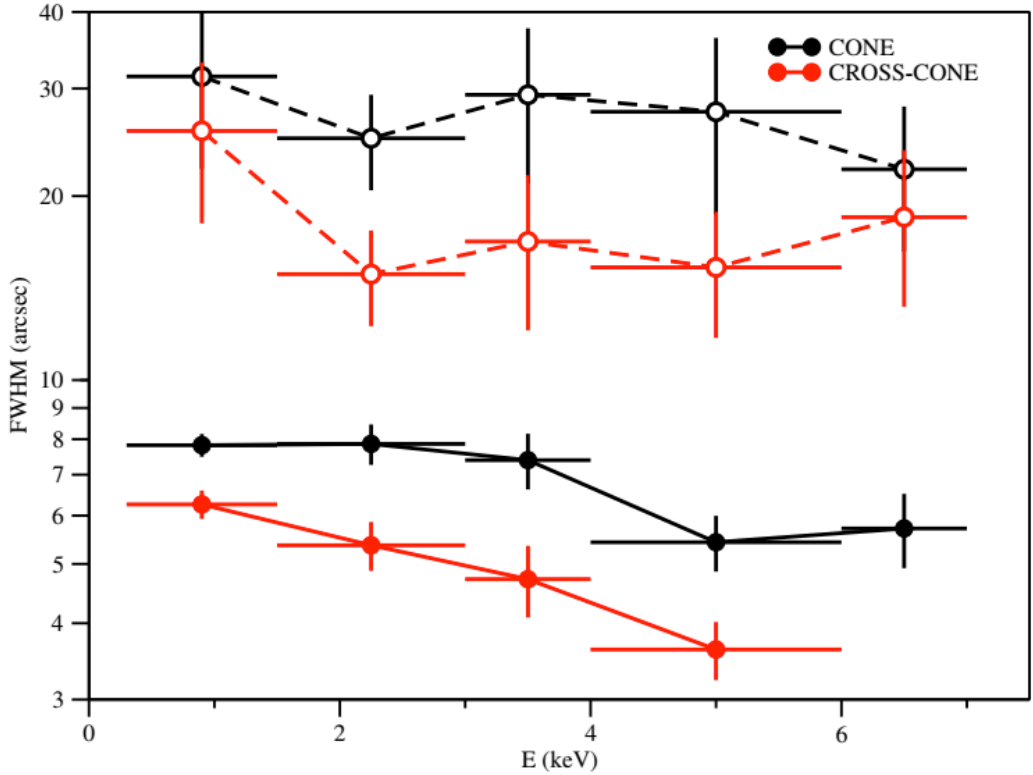

Figure 10 – Filled circles: FWHM of the log-scale radial profiles of Figs. 8-11 as a function of energy. The FWHM corresponds to a factor of 100 decrease from peak surface brightness. Black: cone profiles. Red: cross cone profiles. Empty circles: Full-Width at the level where the background-subtracted surface brightness is consistent with the field background in the same energy range. The errors are ~1σ.

Extended emission in the cross-cone sectors is at odds with the standard picture of AGNs heavily obscured by a circumnuclear 'torus' that would allow radiation to escape only in the cone direction (the standard model, see Netzer 2015 and references therein). This cross-cone emission may suggest that the torus is not perfectly obscuring. A similar conclusion was reached in the case of NGC 4151 (Wang et al 2011c) and Mkn 573 (Paggi et al 2012).

With our deep observations of ESO 428-G014 we have found that the relative amount of cross-cone and cone extended emission is roughly constant at energies>1.5 keV. Below 1.5 keV the cross-cone component becomes relatively more prominent. This is shown in Table 7, which lists the ratio of 'diffuse' counts in cross-cone versus cone sectors from the 1.5" – 8" circumnuclear annulus for different energy bands. Regions containing non-nuclear point sources were excluded. The diffuse counts were derived by subtracting the nuclear PSF contribution for each energy band, using the same PSF normalization as for the radial profiles. As shown by the profiles (Figs. 7-9) a nuclear point-source is only clearly detected above 3 keV. Below 3 keV, the assumed PSF contribution is an upper limit on what such contribution may be; this emission is dominated by the extended component. This is not going to affect our results, since the expected PSF counts are a few and generally similar in similar areas.

In Table 7 we report both the ratio of total net counts in the cross-cone sectors to the total net counts in the cone sectors (see Fig. 1 for the sectors): (NE+SW)/(NW+SE), and the count ratio (NE+SW)/(NW+NW). The latter may be more representative of the emission related to nuclear photons and their interaction with the ISM, because the SE sector includes the high surface brightness feature corresponding to the radio jet, which may enhance the X-ray emission (Paper I). No significant excess is detected in the cross-cone direction (Paper I) in the 6.1-6.6 keV band (Fe K$\alpha$), so we omit this from the ratios. This ratio would be consistent with 0.3 at 1$\sigma$, but the uncertainties are large.

Table 7 – Extended Component "Cross-Cone / Cone" Count Ratios

| Energy band (keV): | 0.3-3. | 0.3-1.5 | 1.5-3. | 3.-6. | 3.-4. | 4.-5. | 5.-6. |
|---|---|---|---|---|---|---|---|
| (NE+SW) / (NW+SE) | | | | | | | |
| Ratio | 0.43 | 0.47 | 0.36 | 0.30 | 0.31 | 0.25 | 0.52 |
| Error | 0.02 | 0.03 | 0.03 | 0.06 | 0.07 | 0.10 | 0.21 |
| (NE+SW) / (NW+NW) | | | | | | | |
| Ratio | 0.70 | 0.84 | 0.49 | 0.50 | 0.50 | 0.51 | 0.49 |
| Error | 0.02 | 0.06 | 0.04 | 0.08 | 0.10 | 0.16 | 0.25 |

## 5. Discussion

It is generally accepted that AGNs are powered by accretion onto massive nuclear black holes (Rees 1984), and that the variety of observed properties over the emission spectrum can be explained with different amounts of obscuration of the nuclear region together with our line of sight to the AGN ('standard' model, see review, Netzer 2015). AGNs also provide a key ingredient into the evolution of galaxies by interacting with their hosts via radiation, relativistic jets, and slow winds and influencing star formation (e.g. Sazonov, Ostriker & Sunyaev, 2004).

Deep high-resolution observations of nearby AGNs with *Chandra* have provided new evidence that constrains further the AGN model and that allows us to study AGN-galaxy interactions. *Chandra* imaging/spectral studies from a few kiloparsecs down to the inner ~100pc from the AGN in nearby Seyfert galaxies (e.g., NGC 4151, Wang et al 2009, 2010, 2011a, b, c, Mkn 573, Paggi et al 2012) have suggested that the 'standard' model needs to be modified to include leaky absorbers, and have discriminated between clouds shock-heated by nuclear radio jets and those photoionized from the AGN radiations. Our study of ESO 428-G014 continues this line of investigation.

Below we discuss the implications of our analysis for the composition of the AGN emission, and for constraining the truly nuclear component (Section 5.1). We explore the possible physical states of the circumnuclear region, the region where the AGN interacts with the host galaxy ISM, based on the constraints posed by the spectral analysis (Section 5.2). The spatial and spectral characteristics of the kiloparsec-extended emission set constraints on the composition of the ISM of the host galaxy (Section 5.3), and on the physical properties of the circumnuclear absorber (the 'torus' of the standard model; Section 5.4). Finally, our data show the presence of soft 'quasi-spherical' emission from the ISM, suggesting gas thermalized in the host galaxy potential, possibly unconnected to the ionization cone (Section 5.5).

## 5.1 Extended Line and Continuum Components – Luminosities and Inferred AGN Luminosity

The 0.3 - 8.0 keV X-ray spectrum of the nuclear and circumnuclear emission of ESO428-G014 is characterized by a flat continuum component and Fe Kα emission lines at energies >3.0 keV, and by several, partially confused, emission lines in the lower energies (<3.0 keV; Fig. 2, Table 2). Our deep cumulative exposure allows us not only to fit the entire emission, but also to separate the inner circumnuclear region within 1.5'' (~170 pc) radius from the more extended emission (8" ~900 pc outer radius). Using the same empirical model as Levenson et al (2006) for the soft, emission-line dominated spectrum, we find that the overall spectrum (from the 8'' circle) is well fit with a PEXRAV nuclear reflection component with Γ=1.8 dominating at the higher energies > 3 keV, plus an additional power-law continuum with Γ=1.8 (that could be due to unabsorbed AGN photons escaping in the direction of the axis of the torus – the ionization cone direction), and a set of blended Gaussian emission lines (Table 2; resulting from the interaction of AGN photons with the galaxy ISM) dominating at the lower energies. Table 8 summarizes the luminosities of the different spectral components.

Table 8.
Luminosity ($L_X/10^{40}$ erg s$^{-1}$) in Spectral Components and $L_{AGN}$

| Region | $L_X$(Lines)[a] 0.3-3. keV | $L_X$(PEXRAV)[b] 0.3-3.0 keV 3.0-7.0 keV | $L_X$(Γ=1.8)[a] 0.3-3.0 keV 3.0-7.0 keV | $L_X$(FeK) I XXV | $L_{AGN}$(Fit)[c] | $L_{AGN}$(Fe)[d] | $L_{AGN}$(OIII)[e] | $L_{AGN}$(IR)[f] |
|---|---|---|---|---|---|---|---|---|
| ≤1.5" (170pc) | 0.47(1.01) | 0.1 (0.12/41.2) 0.9 (0.94/17.5) | 0.30 (0.54) 0.23 (0.23) | 0.55 0.12 | 60.2 | 300 | 760 | 4100 |
| 1.5"-8" (170pc-900pc) | 0.22(0.46) | 0.01 (0.01/4.5) 0.10 (1.0/1.9) | 0.27 (0.48) 0.23 (0.23) | 0.12 --- | | | | |

a) Observed (Unabsorbed) – Unabsorbed is corrected for line of sight $N_H$ only.
b) Observed (Unabsorbed/Incident) - Incident is the nuclear luminosity incident on the absorber/reflector.
c) $L_{AGN}$(Fit) is estimated from the total intrinsic PEXRAV and power-law contribution from our fit in the 0.3-7.0 keV range.
d) $L_{AGN}$(Fe) is estimated from our Fe Kα I line luminosity of Levenson et al (2006) in the 2-10 keV energy range. Our Fe Kα I luminosity in the 1.5'' circle is consistent with Levenson et al (2006).
e) $L_{AGN}$(OIII) is from Levenson et al (2006).
f) $L_{AGN}$(IR) is the 8-1000 μm luminosity from IRAS flux in Levenson et al (2006).

The main difference in the spectral fit from the inner and outer regions is (1) the relatively stronger PEXRAV component in the inner region, and (2) the relatively stronger $\Gamma=1.8$ power-law in the outer region, to fit the overall steeper continuum spectrum. In the outer region, the PEXRAV component is needed to fit the Fe K$\alpha$ line. As remarked in Paper I, the relatively low ionization parameters far away from the nucleus are consistent with a Compton scattering plus fluorescence origin for the large ~2 keV Fe K$\alpha$ observed EW. In Paper I we argued that this steep continuum is unlikely to originate from undetected X-ray binaries in the emitting region, given the morphology of the emission that follows the extended soft emission, and also because of the corresponding spatially extended Fe K$\alpha$ emission (see Section 4), with an equivalent width greatly in excess of what is seen from X-ray binaries. We suggested a scattered contribution from the intrinsic Seyfert emission, escaping unattenuated in the cone direction since, in the standard unified scheme, there is no screen in this direction. If, as suggested by the XILLVER fit, the scattering/fluorescence model is sufficient, no extra $\Gamma=1.8$ component is needed. However, the presence of soft emission in the cone region at larger radii from the nucleus, where the hard and Fe K$\alpha$ emission are not detected, suggests that soft photons escape the nucleus (Sections 4.3 and 5.3).

Based on our spectral fits, a *substantial fraction of the total observed continuum emission is not associated with the nuclear point source*: in the 1.5''-8'' annulus alone the observed continuum accounts for ~30% of the total 0.3-7.0 keV continuum, but this is a lower limit because the emission within the 1.5'' radius is also partially extended, and the large-scale hard emission extends past 8'' (see radial profiles). The total intrinsic AGN luminosity is in the range $> 6 \times 10^{41} - 4.1 \times 10^{43}$ erg s$^{-1}$, based on different estimates (see also Levenson et al 2006).

**5.2 Implications of the Physical Emission Models**

Our spectral analysis (Section 3) shows that the soft line emission can be adequately fitted only with complex models. In the central 1.5'' radius circumnuclear region, models consisting of 3 photoionization components plus a thermal component with kT=0.4, or alternatively 2 photoionization components and 2 thermal components with kT~0.7 and 1.2 keV, best reproduce the line features over the entire energy range (Table 6, Fig. 3). These models suggest that shocks of at least a few hundred km s$^{-1}$ are present in the central, 170 pc (1.5") radius, circumnuclear region. Interestingly, this is the region that includes the inner radio jet and the braided H$\alpha$ and [OIII] emission line feature, which may be connected to shocks originated by the interaction of the radio jet with the local ISM (Falcke et al 1998). In the extended circumnuclear region (170-900 pc; 1.5" – 8"), thermal only models reach higher temperatures (Table 4), and would imply shock velocities as high as ~ 1800 km s$^{-1}$. Although these high-velocity shocks are not required in the preferred mixed photoionization-thermal models, we point out that they have been reported in CT AGNs both near the nucleus (see the case of NGC6240, Wang et al 2014), and at larger radii (Fischer at al 2013).

Table 9. summarizes the gas parameters we derive from the thermal emission components in the mixed model consisting of 3 photoionization +1 thermal components,

assuming a spherical volume of 1.5'' radius for the inner nuclear region, and 1/10 of the spherical annulus extending to 8'' for the extended emission, to take somewhat into account the morphology of the emission. Both are likely to overestimate the volume filled by a hot gas, given that the emission is rather lumpy.

Table 9.
Thermal Luminosity (0.3-8keV $L_X$) and Thermal Gas Parameters

| | Model: 3Phothoionization+1Thermal | | | | | | |
|---|---|---|---|---|---|---|---|
| Region | kT (keV) | $L_X$ (erg s$^{-1}$) | V (cm$^3$) | $n_e$ (cm$^{-3}$) | $M_{gas}$ ($M_\odot$) | $E_{th}$ (erg) | $\tau_{cool}$ (yr) |
| ≤1.5" (≤170 pc) | 0.4 | $2.8\times10^{39}$ | $5.9\times10^{62}$ | 4 | $2\times10^6$ | $4.3\times10^{54}$ | $4.7\times10^7$ |
| 1.5"-8" (~170- 900 pc) | 0.7 | $1.7\times10^{39}$ | $8.9\times10^{63}$ | 0.7 | $5\times10^7$ | $2.1\times10^{55}$ | $3.9\times10^9$ |

The densities we derive from both thermal and photoionization components (Table 9, see Table 6 for photoionization) are consistent with typical ISM densities, suggesting that these models may indeed represent the physical reality of the emission regions. For the thermal component in the inner 1.5'' radius region, the thermal energy is $\sim 4\times10^{54}$ erg. Given the cooling time of this gas, and assuming efficient heating processes, a nuclear energy input of $\sim10^{44}$ erg s$^{-1}$ could be responsible for the heating. Alternatively, only a few 1000 SNae could be responsible for this heating (given the cooling time this would correspond to a low SN rate $\sim10^{-4}$ yr).

The constraints on the ionization parameter, log U, from the best-fit photoionization components (Table 6) can be compared with the values found for X-ray warm absorbers (WAs). WAs are signatures of highly ionized gas outflowing from the central continuum source, with line-of-sight velocities of 1000 – 2000 km s$^{-1}$, comparable to the velocity seen in the AGN bi-cones (Fischer et al 2013). The column densities in these WAs lie in the $10^{20} – 10^{21}$ cm$^{-2}$ range, as in ESO 428-G014.

Identifying emission spectral components with the WA absorption components would give the covering factor of the WAs. This would be important for physical models of the outflowing gas seen as WAs, and is needed to measure the kinetic power carried in these winds (Arav et al. 2013). In ESO 428-G014, if an identification of the photoionized extended emission with WAs is to be made, the inner region (1.5'' radius) is preferred, since WAs have typical distances from 0.01 – 100 pc from the nuclear continuum emission (Krongold et al. 2007; Arav et al. 2015). In the 2 phase (high and low ionization) WA models, NGC 5548 (Andrade-Velazquez et al., 2010) has one velocity component with log $U_{low}$ ~-0.5 and log $U_{high}$ ~+0.7. In NGC 3783 Krongold et al. (2003) find log $U_{low}$ ~-0.8 and log $U_{high}$ ~+0.8, while Kaspi et al. (2002) find log $U_{low}$ ~-1.8 and log $U_{high}$ ~+0.8 in the same object. The best match to these WAs with the ESO 428-G014 photo-ionized emission components is the 2-photoionization + 2-thermal component model: log $U_{low}$ ~-0.9 and log $U_{high}$ ~+0.8 for the inner 1.5'' radius region (Table 6). An identification of the Was with the resolved bi-cone emission in AGN is thus promising, though tentative.

**5.3 Structure of the ISM from the Energy-Dependence of the kpc-Scale Emission**

As shown in Section 4, extended 2-3 kpc-scale emission is detected in the direction of the major axis of ESO 428-G140, which is also the direction of the ionization cone (Falcke 1996, 1998). Fig. 12 shows an energy-related trend in the size of the extended emission, with the softer emission being the most extended. Given that we are comparing FWHM in Fig. 12, this conclusion is not the result of relatively more photons being detected in the softer bands.

In Paper I we have discussed how the extended hard emission (energies >3 keV) may be related to scattering of the AGN emission by interstellar clouds in the disk of ESO 428-G140, although some of the continuum emission may also be related to the presence of a radio jet (we will return to this in Paper III). Here we extend these considerations to include the entire *Chandra* energy band.

For us to be able to detect the AGN emission propagating in the cone direction, we need these photons to interact with the ISM. Our spectral results show that both photoionization and collisional interactions are possible in the <3 keV emission-line dominated spectral band. These require low column density clouds with $N_H \sim 10^{22}$ cm$^{-2}$ (Zombeck 2007), or at least a low column density 'skin' surrounding denser clouds. If the harder diffuse emission is also due to Compton scattering off ISM clouds (Paper I), it will require optically thicker clouds ($N_H > 10^{24.5}$ cm$^{-2}$) to scatter the radiation efficiently given the electron scattering cross-section (e.g. Allen, 1973). In this case the smaller FWHMs we measure in the hard continuum and Fe K$\alpha$ profiles imply that the optically thicker clouds responsible for this scattering may be relatively more prevalent closer to the nucleus. These clouds must not prevent soft ionizing X-rays from the AGN escaping to larger radii, in order to have photoionized ISM at larger radii. Therefore, these scattering clouds must be clumped and so have higher density. This suggests that at smaller radii there may be a larger population of molecular clouds to scatter the hard X-rays, as in the Milky Way (e.g., Nakanishi & Sofue 2006).

**5.4 Leaky Absorber Constraints**

The diffuse circumnuclear X-ray emission of ESO 428-G014 is significantly extended both in the cone (ionization bi-cone and radio jet direction) and in the cross-cone direction (Paper I; Section 4.3, Fig. 12). The extent in the cross-cone direction is at odds with AGN models that require the presence of a uniformly optically thick torus around the AGN nucleus, perpendicular to the bi-cone axis (Stalevski et al. 2016, and refs. therein). A similar effect was reported for NGC 4151 (Wang et al 2011c) and Mkn 573 (Paggi et al 2012), where the suggestion was advanced that the torus lets some radiation escape. This may also be the case in ESO 428-G014.

Given the uncertainties, the ratio between the FWHM of the cone and cross-cone directions (~1.7) is consistent with the ratio between major and minor axis diameters of

ESO 429-G014 (see Section 2). This similarity may suggest that the smaller extent in the cross-cone direction may be at least in part a disk inclination effect. If this were the case, there could be a uniform 'porosity' of the AGN blanketing clouds that would let photons escape uniformly in all directions. However, we can exclude this possibility: at energies >1.5 keV (see Section 5.5 for a discussion of the emission at lower energies, <1.5 keV) we also find (Table 7) that the photons in the low surface brightness component perpendicular to the bi-cone are ~50% of those in the bi-cone direction. Based on the projected angles on the sky, this would imply as a minimum ~30% further reduction of the count rate in the cross-cone direction.

If we assume, as in the standard model, that the absorber is a torus, we can estimate the transmission of the torus in the cross-cone direction, by considering the cone and cross-cone solid angles. Since the cross-cone region is ~5 times larger in volume than that of the bi-cone, the transmission of the obscuring region in the cross-cone direction would be ~10% of that in the cone-direction. The lack of energy dependence (above 1.5 keV) of the ratio of photons escaping from the cross cone and cone regions is in agreement with the partial obscuration picture, in which a fraction of photons escape in the cross-cone region, with similar spectral distribution as those (the larger fraction) escaping in the cone region. The interaction of the escaping photons with the ISM clouds would cause the extended low-surface brightness diffuse emission, in both cone and cross-cone directions.

Models for the AGN "obscuring torus" require clumping for the dust to survive (Krolik & Begelman 1985), and also to allow outer clouds to receive sufficient illumination to re-emit the observed levels of mid-IR emission (Nenkova, Ivezic & Elitzur 2002). Clumpy tori seem consistent with the porous absorber required in ESO428-G014 and other AGN. However, models of clumpy tori have become steadily more sophisticated (e.g. Garcia-Gonzalez et al., 2017), and now include a 2-phase medium. The X-ray optical depth through the inter-cloud medium appears to be too large to let any significant fraction of keV X-rays escape (Stalevski et al., 2016; Stalevski 2017, private communication). Perhaps some larger-scale clumping without an inter-clump medium is also present?

We also see an enhancement of the surface brightness of the extended emission along the direction of the radio jet and ionization cone (Paper I; Figs. 4-5) in the inner ~5'' (~850 pc) region, especially in the SE direction. This suggests an additional preferential escape direction from the nuclear region, and then additional photoionization and collisional ionization along this preferential cone axis. This region will be explored fully in Paper III.

**5.5 A thermalized 'quasi-spherical' soft gaseous component**

In the 0.3-1.5 keV band, we find a significantly larger ratio of cross-cone to cone detected counts (~84 %, excluding possible jet effects in the SE sector; Table 7). In this spectral band, the emission is also rounder than at higher energies (see comparisons with the 1.5-3.0 keV band in Figs. 4, 7, and 10). This excess of soft photons cannot be explained in the AGN torus obscuration picture. It requires a different source for a good amount of this soft diffuse emission. A natural candidate is a hot ISM partially trapped in the potential of the galaxy (David et al 2006). If we assume that the cross-cone counts due to the AGN are 50% of the cone counts, as seen for higher energies (Table 7), then we find an excess 173 cone counts. For a thermal APEC spectrum with kT~0.7 keV, the corresponding luminosity (in the *Chandra* observable band, > 0.3 keV) would be ~5×10$^{38}$ erg s$^{-1}$, mostly detectable below 1.5 keV. This luminosity can account for a substantial fraction of the thermal component suggested by the spectral fits in the 1.5''-8'' annulus (Table 9). Interestingly, this luminosity is similar to that detected in the ~5 kpc extended component detected in the deep *Chandra* observations of NGC 4278, a gas-poor elliptical hosting a compact radio source, where interaction with the AGN may be responsible for heating the ISM (Pellegrini et al 2012). We cannot exclude that supernova heating is also responsible. Given the ~10$^9$ yr cooling time of this component and the ~10$^{55}$ erg energy content (Table 9), the heating would require a supernova rate of only ~10$^{-5}$ yr$^{-1}$.

## 6. Summary and Conclusions

We have analyzed the deep *Chandra* observation (~155 ks) of the CT AGN ESO428-G014 to study the kpc-scale diffuse X-ray emission both spectrally and spatially as function of energy. In summary, we find that:

1) Confirming the early results of Levenson et al (2006), the 0.3 - 7.0 keV X-ray spectrum is characterized by a soft continuum at energies < 3.0 keV, which shows the signature of line emission; a hard power-law continuum plus the 6.4 keV Fe Kα line at energies > 3 keV.

2) As reported by Levenson et al (2006) for the soft component, and by ourselves in Paper I, all these spectral components are associated with large-scale spatial components, although point-like nuclear emission can also be seen at the higher energies and in the Fe Kα line. The observed continuum emission in the 1.5''-8'' (~170 – 900 pc) annulus accounts for ~30% of total observed continuum in the 0.3-7.0 keV band. In the hard band (3.0-7.0 keV), 22% of the continuum is observed in the 1.5''-8'' annulus.

3) Using a nuclear reflection model + additional power-law and a set of Gaussian lines (as in Levenson et al 2006), we analyzed separately the emission in the region within ~170 pc of the AGN, and in a surrounding ~170-900 pc annulus. This model fit suggests the presence of blended OVII, OVIII, NeIX, NeX, and several Fe lines at energies <1.2 keV. We also detected more isolated emission from lines that can be identified with Mg (XI, XII), Si Kα, and S Kα. The main differences in the spectral fit from the inner and outer regions, are the relatively stronger reflection component in the inner region, and a relatively stronger Γ=1.8 power-law in the outer region, which could be due to unprocessed AGN photons escaping along the axis of the torus. The Fe Kα emission is dominated by the neutral 6.4 keV line. Fe XXV accounts for ~20% or less of the line emission in the 6-7 keV band.

4) Analyzing the spectrum with photoionization and thermal codes, we find that the models that best fit the spectrum at all wavelengths are composite photoionization + thermal models (either 3 photoionization + 1 thermal component or 2 photoionization + 2 thermal components), especially in the central 1.5'' (170 pc) region. The densities derived from both thermal and photoionization models are consistent with typical ISM densities of a few atoms per cubic centimeter or less. In the 3 photoionization + 1 thermal model, the thermal component has kT ~ 0.4 keV in the inner 170 pc region (with cooling time ~$10^7$ yrs), and kT ~0.7 keV in the surrounding annulus (170-900 pc, with cooling time ~$10^9$ yrs). Either heating from the AGN or supernova explosions can explain the thermal energy content of this gas.

5) The photo-ionization components with constrained values of ionization parameter, log U, in the central 1.5" (170 pc) circle compare reasonably well with the values found for X-ray warm absorbers (WAs). The best match to WAs detected in other Seyfert galaxies with the ESO428-G014 photo-ionized emission components is the 2-component model: log $U_{low}$ ~-0.9 and log $U_{high}$ ~+0.8 for the inner 1.5" radius region.

6) We detect extended 2-3 kpc-scale (18"-26") emission in the direction of the major axis of ESO428-G140, which is also the direction of the ionization cone (Falcke et al. 1996; see Section 4). We find an energy-related trend in the size of the extended emission, with the softer emission (<3 keV) being more extended than the harder emission. The smaller extent of the hard continuum and Fe Kα profiles imply that the optically thicker clouds responsible for this scattering may be relatively more prevalent closer to the nucleus. These clouds must not prevent soft ionizing X-rays from the AGN escaping to larger radii, in order to have photoionized ISM at larger radii. Therefore, these scattering clouds must be clumped and so have higher density. This suggests that at smaller radii there may be a larger population of molecular clouds to scatter the hard X-rays, as in the Milky Way (e.g., Nakanishi & Sofue 2006).

7) The diffuse emission is also significantly extended in the cross-cone direction, where the AGN emission would be mostly obscured by the torus in the standard AGN model). Excluding the SE region affected by interaction with the radio jet, the ratio of detected counts (i. e. photons) in the cross-cone to the cone region is constant to ~50% above 1.5 keV. This and the spatial properties of the emission suggest that in the standard model the transmission of the obscuring torus in the cross-cone direction is ~10% of that in the in-cone direction. The lack of energy dependence (above 1.5 keV) of the ratio of photons escaping from the cross cone and cone regions is in agreement with the partial obscuration picture, in which a fraction of photons escape in the cross-cone region, with similar spectral distribution as those escaping in the cone region.

8) In the 0.3-1.5 keV band, the ratio of cross-cone to cone photons increases to ~84%, suggesting an additional soft emission component, disjoint from the AGN. The emission is also rounder then at higher energies, with relatively closer FWHM for the cone and cross cone regions. This softer component could be due to a hot ISM trapped in the potential of the galaxy (see David et al 2006). The luminosity of this component ~$5 \times 10^{38}$ erg s$^{-1}$, is a large fraction of the thermal component suggested by our spectral fits in the 170-900 pc annulus. This luminosity is similar to that detected in the ~5 kpc extended component detected in the deep *Chandra* observations of NGC 4278, a gas-poor elliptical hosting a compact radio source, where interaction with the AGN may be responsible for heating the ISM (Pellegrini et al 2012).

In conclusion, this paper shows how the AGN-galaxy interaction may shape the X-ray emission of AGNs at all energies and in a wide range of spatial scales, giving information both on the host ISM and on the AGN itself. The deep *Chandra* observation of ESO 428-G014 demonstrate how high-resolution spatial/spectral X-ray data, with good statistical significance, can provide surprising results on these 'standard' sources. This highlights the scientific importance of a future high sub-arcsecond resolution X-ray telescope, with much larger collecting area than *Chandra*, such as *Lynx*[5].

We retrieved data from the NASA-IPAC Extragalactic Database (NED), and the *Chandra* Data Archive. For the data analysis we used the *CIAO toolbox*, *Sherpa* and *DS9*, developed by the Chandra X-ray Center (CXC); and *XSPEC* developed by the HEASARC at NASA-GSFC. This work was partially supported by the *Chandra* Guest Observer program grant GO5-16090X (PI: Fabbiano), and by NASA contract NAS8-03060 (CXC). The work of J.Wang was supported by the National Key R&D Program of China (2016YFA0400702) and the National Science Foundation of China (11473021,11522323).

---

[5] https://wwwastro.msfc.nasa.gov/lynx/

# Appendix A.

## Additional Image and Radial Profiles

For completeness we include here (Figs. A.1) the image of the entire soft emission line-dominated spectral component (0.3-3.0 keV). This image was analyzed as explained in Section 4.1, and additionally slightly adaptively smoothed. The color scale was stretched to emphasize the lowest surface brightness emission. This results in some loss of the details visible in the figures in the main text. The corresponding radial profiles are shown in Fig. A.2 together with those of the entire 3.0-6.0 keV hard continuum. The 3.0-6.0 keV hard continuum image was published in Paper I.

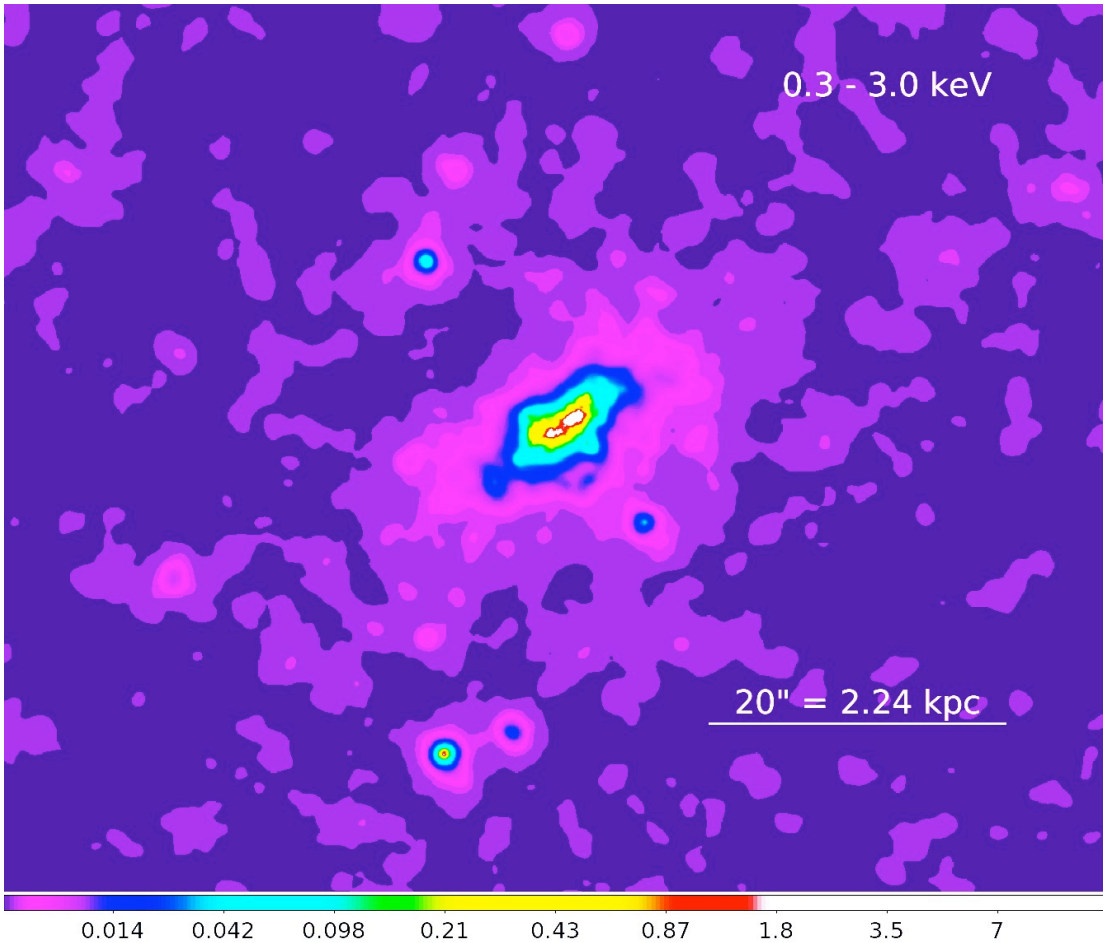

Fig. A.1 – EMC2 PSF deconvolution image in the 0.3-3.0 keV band (se text for details). Color scale is in number of counts per image pixel. N is to the top and E to the left.

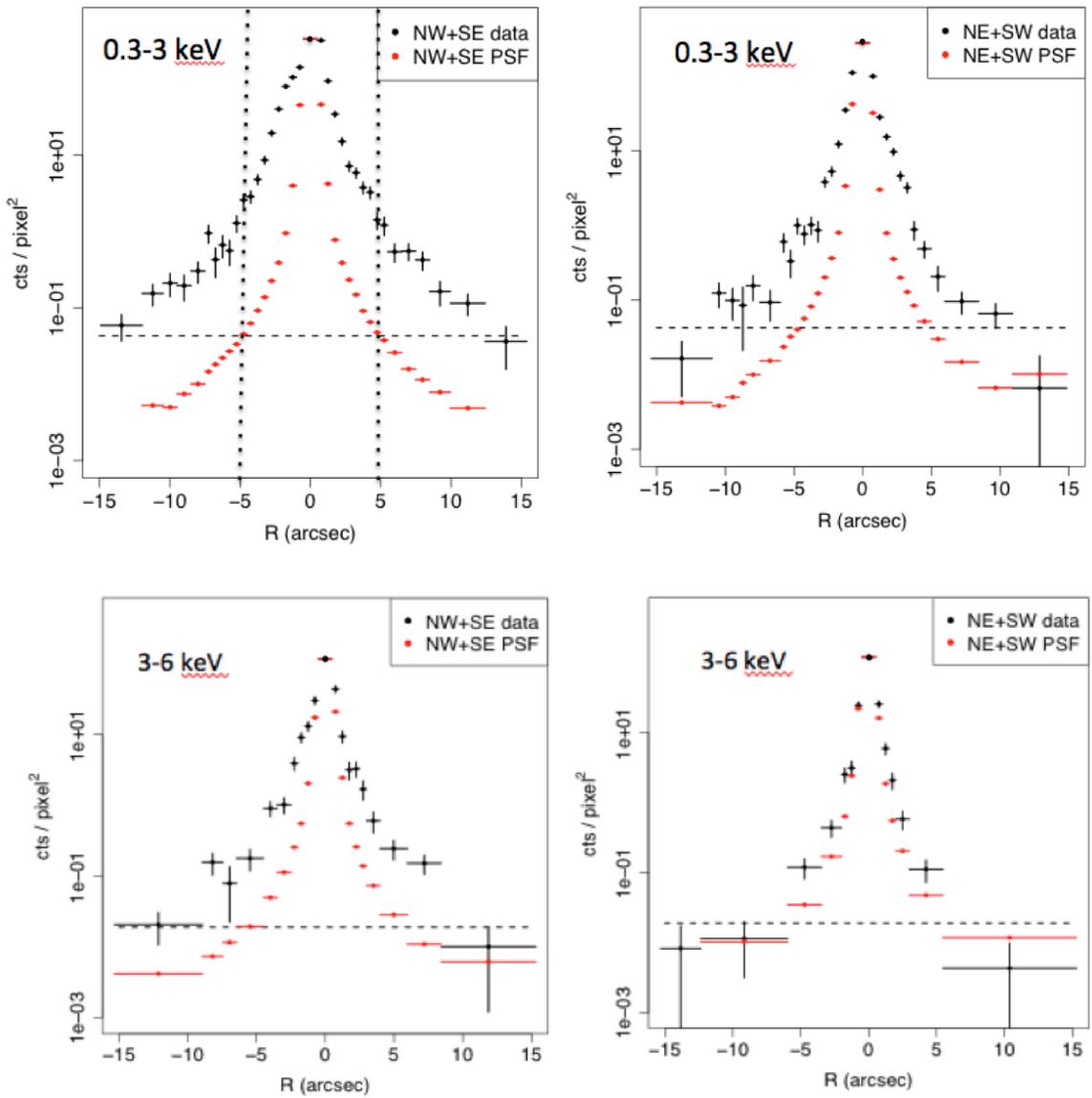

Fig. A.2 – See caption of Fig. 7. The vertical dashed lines in the 0.3-3.0 keV cone profile emphasize the inner 5" region, where the surface brightness is larger in the SE compared to the NW direction. Past 5" the SE and NW profiles are comparable.

Allen, C.W., 1973, Astrophysical Quantities, 3rd ed. [Athlone Press:London], p.93.

Andrade-Velázquez, M., Krongold, Y., Elvis, M., Nicastro, F., Brickhouse, N., Binette, L., Mathur, S., Jiménez-Bailón, E. 2010, ApJ, 711, 888.

Arav, N., Borguet, B., Chamberlain, C., Edmonds, D., and Danforth, C., 2013, MNRAS, 4236, 3286.

Arav, N., Chamberlain, C., Kriss, G.A., Kaastra, J.S., Cappi, M., et al., 2015, A&A, 5576, 37.

Arnaud, K. A. 1996, Astronomical Data Analysis Software and Systems V, A.S.P. Conference Series, Vol. 101, 1996, George H. Jacoby and Jeannette Barnes, eds., p. 17.

Bianchi, S., Guainazzi, M., and Chiaberge, M. 2006 A&A, 448, 499.

David, L. P., Jones, C., Forman, W., Vargas, I. M., and Nulsen, P. 2006, ApJ, 653, 207.

Esch, D. N., Connors, A., Karovska, M. and van Dyk, D. A. 2004, ApJ, 610, 1213.

Fabbiano, G., Elvis, M., Paggi, A., Karovska, M., Maksym, W. P., Raymond, J., Risaliti, G., Wang, Junfeng, 2017, ApJL, 842L, 4.

Falcke, H., Wilson, A. S. and Simpson, C. and Bower, G. A. 1996, ApJ, 464, L31.

Falcke, H., Wilson, A. S. and Simpson, C. 1998, ApJ, 502, 199.

Ferland, G. J., Korista, K. T., Verner, D. A., Ferguson, J. W. Kingdon, J. B., Verner, E. M. 1998, PASP, 110, 761.

Fischer, T. C., Crenshaw, D. M., Kraemer, S. B., and Schmitt, H. R.
2013, ApJS, 209, 1.

Foster, A. R.; Ji, L.; Smith, R. K.; Brickhouse, N. S., 2012, ApJ, 756, 128.

Garcia-Gonzalez. J., Alonso-Herrero, A., Hoenig, S.F., Herna'n-Caballero, A., Ramos-Aleida, C., et al., 2017, MNRAS, 470, 2578.

García, J., Dauser, T., Reynolds, C. S., Kallman, T. R., McClintock, J. E., Wilms, J., Eikmann, W. 2013, ApJ, 768, 146.

Karovska, M., Schlegel, E., Hack, W., Raymond, J. C. and Wood, B. E. 2005 ApJ, 623, L137.

Karovska, M., Carilli, C. L., Raymond, J. C. and Mattei, J. A. 2007, ApJ, 661, 1048.

Kaspi, S., Brandt, W. N., George, I. M., Netzer, H., Crenshaw, D. M., Gabel, J. R., Hamann, F. W., Kaiser, M. E., Koratkar, A., Kraemer, S. B., et al. 2002, ApJ, 574, 643.

Krolik J. H., Begelman M. C., 1988, ApJ, 329, 702.

Krongold, Y., Nicastro, F., Elvis, M., Brickhouse, N., Binette, L., Mathur, S., and Jimenez-Bailon, E., 2007, ApJ, 659, 1022.

Levenson, N. A., Heckman, T. M., Krolik, J. H., Weaver, K. A., and Życki, P. T. 2006, ApJ, 648, 111.

Magdziarz, P. and Zdziarski, A. A. 1995, MNRAS, 273, 837.

Maiolino, R., and Riecke, G., 1995, ApJ, 454, 95.

Maiolino, R., Salvati, M., Bassani, L., Dadina, M., della Ceca, R., Matt, G., Risaliti, G., and Zamorani, G. 1998, A&A, 338, 781.

Marinucci, A., Risaliti, G., Wang, J., Nardini, E., Elvis, M., Fabbiano, G., Bianchi, S. and Matt, G. 2012, MNRAS, 423, L6.

Marinucci, A., Bianchi, S., Fabbiano, G., Matt, G., Risaliti, G., Nardini, E., and Wang, J. 2017, MNRAS, 470, 4039.

Nakanishi, H., Sofue, Y., 2006, PASJ, 58, 847.

Nenkova M., Ivezic Z., Elitzur M., 2002, ApJ, 570, L9.

Netzer, H., 2015, ARA&A, 53, 365.

Ogle, P. M., Marshall, H. L., Lee, J. C., Canizares, C. R. 2000, ApJ, 545, 81.

Paggi, A., Wang, J., Fabbiano, G., Elvis and M., and Karovska, M. 2012, ApJ, 756, 39.

Pellegrini, S., Wang, J., Fabbiano, G., Kim, D.-W., Brassington, N. J., Gallagher, J. S., Trinchieri, G., and Zezas, A. 2012, ApJ, 758, 94.

Rees, M. J., 1984, 22, 471.

Risaliti, G., Maiolino, and R. Salvati, M. 1999, ApJ. 522, 157.

Sazonov, S. Y., Ostriker, J. P., and Sunyaev, R. A., 2004, MNRAS, 347, 144

Stalevski, M., Ricci, C., Ueda, Y., Lira, P., Fritz, J., and Baes, M., 2016, MNRAS, 458, 228.

Ulvestad, J. S. and Wilson, A. S. 1989, ApJ, 343, 659.

Wang, J., Fabbiano, G., Karovska, M., Elvis, M., Risaliti, G., Zezas, A., and Mundell, C. G. 2009, ApJ, 704, 1195.

Wang, J., Fabbiano, G., Risaliti, G., Elvis, M., Mundell, C. G., Dumas, G., Schinnerer, E., and Zezas, A. 2010, ApJ, 719, L208.

Wang, J., Fabbiano, G., Risaliti, G., Elvis, M., Karovska, M., Zezas, A., Mundell, C. G., Dumas, G. and Schinnerer, E. 2011a, ApJ, 729, 75.

Wang, J., Fabbiano, Elvis, M., G., Risaliti, Mundell, C. G., G., Karovska, M., and Zezas, A., 2011b, ApJ, 736, 62.

Wang, J., Fabbiano, G., Elvis, M., Risaliti, G., Karovska, M., Zezas, A., Mundell, C. G., Dumas, G. and Schinnerer, E. 2011c, ApJ, 742, 23.

Wang, J., Nardini, E., Fabbiano, G., Karovska, M., Elvis, M., Pellegrini, S., Max, C., Risaliti, G., U, V., and Zezas, A. 2014, ApJ, 781, 55.

Zombeck M.V, (2007 "Handbook of Space Astronomy and Astrophysics", [CUP: Cambridge], p.280